\documentclass{aa}  

\usepackage{graphicx}
\usepackage{txfonts}
\usepackage{lipsum}
\usepackage{subcaption}         
\usepackage{lscape}             
\usepackage{placeins}           
                                
\usepackage[colorlinks=true, linkcolor=blue, citecolor=blue, urlcolor=blue]{hyperref}

\usepackage{graphicx}	
\usepackage{amsmath}	

\usepackage{lipsum}
\usepackage[english]{babel}
\usepackage{color}
\usepackage[dvipsnames]{xcolor}

\usepackage{cleveref}
\usepackage{siunitx}
\usepackage{longtable}
\usepackage{comment}
\usepackage{booktabs}

\begin{document}

   \title{V407 Vul: a triple star system with an AM CVn detectable by gravitational wave observatories}


%
%
%

   \author{James Munday\inst{1,2}\corrauth{james.munday98@gmail.com}        
    \and Antonio C.\ Rodriguez\inst{3,4}
    \and Nicole Reindl\inst{5}
    \and Nina Mackensen\inst{5}
    \and Tin Long Sunny Wong\inst{6}
    \and S.P.\ Littlefair\inst{7}
    \and Ingrid Pelisoli\inst{2}
    \and Alex Brown\inst{8}
    \and N. Castro Segura\inst{2}
    \and Joheen Chakraborty\inst{9}
    \and Harry Dawson\inst{1}
    \and V. S.\ Dhillon\inst{7,10}
    \and Matti Dorsch\inst{1}
    \and Martin Dyer\inst{7,11}
    \and James A. Garbutt\inst{7}
    \and Stephan Geier\inst{1}
    \and Matthew Green\inst{12,13}
    \and Dan Jarvis\inst{5}
    \and Mark R. Kennedy\inst{14}
    \and Paul Kerry\inst{7}
    \and James McCormac\inst{2}
    \and Steven G. Parsons\inst{7}
    \and Jan van Roestel\inst{15}
    \and Dave Sahman\inst{7}
    \and P.-E.\ Tremblay\inst{2}
    \and Amalie Yates\inst{7}
        }

\institute{Institute for Physics and Astronomy, University of Potsdam, Karl-Liebknecht-Str. 24/25, 14476 Potsdam, Germany
\and Department of Physics, Gibbet Hill Road, University of Warwick, Coventry CV4 7AL, United Kingdom
\and Center for Astrophysics | Harvard \& Smithsonian, 60 Garden St, Cambridge, MA, 02138, USA
\and Department of Astronomy, California Institute of Technology, 1200 East California Blvd, Pasadena, CA, 91125, USA
\and Zentrum für Astronomie der Universität Heidelberg, Landessternwarte, Königstuhl 12, D-69117 Heidelberg, Germany
\and The Observatories of the Carnegie Institution for Science, Pasadena, CA 91101, USA
\and Astrophysics Research Cluster, School of Mathematical and Physical Sciences, University of Sheffield, Sheffield S3 7RH, UK
\and Hamburger Sternwarte, University of Hamburg, Gojenbergsweg 112, 21029 Hamburg, Germany
\and Department of Physics \& Kavli Institute for Astrophysics and Space Research, Massachusetts Institute of Technology, Cambridge, MA 02139, USA
\and Instituto de Astrof\'isica de Canarias, E-38205 La Laguna, Tenerife, Spain
\and Research Software Engineering, University of Sheffield, Sheffield, S1 4DP, UK
\and Homer L. Dodge Department of Physics and Astronomy, University of Oklahoma, 440 W. Brooks Street, Norman, OK 73019, USA
\and JILA, University of Colorado and National Institute of Standards and Technology, 440 UCB, Boulder, CO 80309-0440, USA
\and School of Physics, University College Cork, Cork, T12 K8AF, Ireland
\and Institute of Science and Technology Austria, Am Campus 1, 3400, Klosterneuburg, Austria}


   \date{Received June 18, 2026}

 
  \abstract
   {The AM~CVn class includes mass transferring, ultra-compact double white dwarf binaries with orbital periods on the timescale of minutes. 
A long-standing puzzle is that none of the roughly fifty ultra-compact, ``verification binaries'' which are easily detectable in the millihertz gravitational wave regime reside in a triple star configuration. Much evidence has hinted at V407~Vul being an inspiraling, double white dwarf AM~CVn with an orbital period of 569\,s. Yet, a decisive confirmation has proved challenging since a main sequence star dominates its visible spectrum. We present a clear confirmation of the triple star nature of the source by detecting a significant astrometric wobble of the photocentre on the 569\,s orbital period of the binary. The AM~CVn and the main sequence components are gravitationally bound with a spatial separation of roughly 0.03--0.04\arcsec, equating to an orbital separation of approximately 120\,AU. A total of 23 years of orbital timing constrained the orbital decay of the AM~CVn as being precise to the 1\% level, critical in understanding if this class of binary survives through a period minimum or coalesce. New Hubble Space Telescope ultra-violet imaging and spectroscopic data allowed the isolated detection of the AM~CVn at shorter wavelengths, revealing an approximately 58\,000\,K accretor white dwarf, while placing a firm distance constraint of $3510^{+140}_{-110}$\,pc. At this distance, we predict that the Laser Interferometer Space Antenna (LISA) will detect V407~Vul with a $28.4\pm9.2$ signal-to-noise ratio in a 4\,yr mission time, making it the first verification binary with an outer tertiary, or ``verification triple'', detectable for millihertz gravitational wave observatories.}

   \keywords{stars: close binaries -- accretion -- gravitational waves
               }
   \maketitle
   \nolinenumbers

\section{Introduction}
AM~CVn binary star systems are ultra-compact (P$_{\mathrm{orb}}\approx5-65$\,min), semi-detached systems in which a white dwarf (WD) accretes from a hydrogen-deficient donor. Such systems originate from three donor evolutionary paths: 1) The double WD (DWD) channel \citep{Paczynski1967, Webbink1984, Deloye2007thermalEvolutionAMCVn}. 2) The helium star channel, consisting of a non-degenerate or semi-degenerate, helium-rich/helium burning donor \citep{Savonije1986, Iben1987}. 3) The evolved cataclysmic variable channel with an evolved main sequence star \citep{Tutukov1985, Augusteijn1993, Thorstensen2002, Podsiadlowski2003, ElBadry2021evolvedCV}. 
A gradual loss of orbital angular momentum draws the stars closer until the larger radius donor overflows its Roche lobe. The helium star and main sequence channels begin mass transfer with an orbital period up to around 65\,min, exhibiting a high mass transfer rate that increases as the binary approaches an orbital period minimum of around $\approx10-15$\,min \citep[e.g.][]{NelemansII2001populationSynthesisOfWDsAMCVn}. On the other hand, AM~CVns formed through the DWD channel initiate significant mass transfer ($>10^{-10}$\,M$_\odot$\,yr$^{-1}$) at $\approx10$--$20$\,min owing to the much more compact nature of the donor. The Roche lobe of the donor is filled at a smaller orbital separation, maintaining a $\lessapprox10^{-8}$\,M$_\odot$\,yr$^{-1}$ mass transfer rate until close approach to the period minimum of $\approx5-10$\,min \citep{Deloye2007thermalEvolutionAMCVn, Kaplan2012OrbitalEvolutionOfCompactWDBinaries, Wong2021AMCVnMESAPureHe}, or to the point of the stars coalescing.

The star systems HM~Cancri \citep{RamsayHakalaCropper2002HMCncBinarydiscoveryMaybe, Israel2002HMCncDiscovery} and V407~Vul \citep{Motch1996, Cropper1998v407vulDiscovery} were first identified through pulsed X-ray and optical signals of period 321\,s and 569\,s, respectively \citep{Ramsay2000v407Vul}. HM~Cancri originates from the DWD channel \citep{Roelofs2010HMCncMassRatio, MundayHMCnc2023}, making it the most-compact AM~CVn and binary star system discovered to date, and the exact nature of V407~Vul has been debated in the last decades. V407~Vul was first thought to be a double degenerate polar \citep{Cropper1998v407vulDiscovery}, but later data and analysis by \citet{Ramsay2002v407VulSpectAndPhot} revealed that a lack of any prominent optical emission lines and a lack of polarisation makes this characterisation unlikely. \citet{Strohmayer2002pdotV407VulXray, Strohmayer2004pdotXray} soon after detected an orbital decay consistent with gravitational wave predictions, suggesting that the periodic signal is related to the orbital period of the binary, as was supported by \citet{Ramsay2005v407Vulxray}. In better accordance with the lack of magnetism while maintaining a compact binary configuration, three other primary theories arose, being those of a DWD binary undergoing direct-impact accretion to cause a bright X-ray luminosity \citep{Marsh2002V407VulDirectImpact}, an intermediate polar model \citep{Norton2004IPmodelHMCnc} or a unipolar inductor \citep{Wu2002UImodel, Wu2009review}. \citet{Barros2005UImodelHMCncV407Vul} showed that no geometrical configuration can justify the arrival phase difference between the X-ray and optical signal, ruling against the unipolar inductor model, and a similar rationale was later used to indicate that a intermediate polar model is improbable too \citep{Barros2007HMCnc}. \citet{Ramsay2008} found that the X-ray spectrum of V407~Vul can be modelled with a single blackbody component so long as there is a significant source enhancement of Neon, while also remarking the clear similarity in the X-ray spectrum of V407~Vul and HM~Cancri \citep[see also][]{Ramsay2005v407Vulxray, Ramsay2006v407Vulxray}. The phase-resolved X-ray pulse profile of HM~Cancri and V407~Vul are very similar too, both reproducible using a 2-component direct-impact model caused by a DWD binary \citep{Wood2009}. Finally, the energy budget of the observed orbital decay is compatible with both the DWD AM~CVn and the unipolar inductor model \citep{Deloye2006predictingFddotFromLacc, DallOsso2007UIHMCnc}. With all the evidence presented and its striking similarity to HM~Cancri as a confirmed such case, V407~Vul appears to most probably be an ultra-compact, DWD AM~CVn binary.

Yet, the double-degenerate scenario has complications due to the fact that a bright G-type star is apparent in the exact location of the pulsating source, dominating in the optical and infrared spectrum \citep{Steeghs2006V407VulGeminiSpectrum}. It is impossible for a G-type star to reside in a 569\,s orbital period binary as it would engulf the WD. There is also a lack of any radial velocity variation from the G-type star, clearly indicating that the G-type star is not one of two members of the close binary \citep{Steeghs2006V407VulGeminiSpectrum}. \citet{Barros2007HMCnc} tentatively claim an astrometric fluctuation away from the pulse origin of approximately 0.027\arcsec\ and that the G-type star could be bound to the binary as a triple star system. For the approximate 1.1--3.5\,kpc distance of V407~Vul \citep{Steeghs2006V407VulGeminiSpectrum}, the orbital period of the triple orbit would be on the timescale of tens to hundreds of years -- far longer than the orbits required for the intermediate polar or unipolar inductor models. This would hence mean that the G-type star has little effect on the architecture of the pulse emitting source. All said, and largely because of the relevance of the G-type star on the system configuration, the true nature of V407~Vul is still not absolute.

In this study, we present and analyse many new spectroscopic and imaging data to solve this conundrum. We also present results that show an isolated detection of the pulse emitting binary in V407~Vul. Section~\ref{sec:Observations} details the new observations taken in this work. In Section~\ref{sec:TimingSolutions} we constrain the shortening of the flux arrival time to a very high precision. Section~\ref{sec:Astrometry} addresses ground-based astrometry as a means to resolve if V407~Vul is a triple. Section~\ref{sec:SEDfittingSpectroscopicPhotometric} involves fitting of new visible-to-near-infrared spectra and Hubble Space Telescope (HST) ultra-violet (UV) photometry combined with all-sky photometry. A discussion of all results involving the spectrophotometric fitting, triple analysis and gravitational wave predictions are discussed in Section~\ref{sec:ResultsAndDiscussion}, before closing with the possible ultimate fates of V407~Vul.


\section{Observations}
\label{sec:Observations}
\subsection{Photometry}
We observed V407 Vul with the high-speed cameras ULTRACAM \citep{ULTRACAM2007}, ULTRASPEC \citep{ULTRASPEC2014} and HiPERCAM \citep{HiPERCAM2016, Hipercam2021Paper}, where ULTRACAM and HiPERCAM observe simultaneously in multiple bands. Our observations began in 2003 and we used the 4.2\,m William Herschel Telescope (WHT), the 3.5\,m ESO New Technology Telescope (NTT) and the 10.4\,m Gran Telescopio Canarias (GTC); a full observing log is supplied in Appendix~\ref{tab:observingLog}. Time-series photometry was also obtained with the 2.5\,m Isaac Newton Telescope (INT) Wide Field Camera. All data were bias-corrected and flat-fielded with the HiPERCAM pipeline. An extra dark correction was applied for ULTRACAM since the instrument runs at a slightly hotter temperature. HiPERCAM $i_s$ and $z_s$ band data were corrected for fringing using pre-obtained fringe maps. A single night of observations with HiPERCAM, phase folded on the orbital period, is shown in Fig.~\ref{fig:timings_hcamLCs}.

We supplemented our own observations with archival data from the WHT Auxiliary Port Imager (API) and the Liverpool Telescope (LT) RATCam. WHT/API data was bias-corrected and flat-fielded with custom scripts, whereas data from LT/RATCam were automatically reduced through the LT's instrument pipeline. We then extracted all data using the HiPERCAM reduction pipeline. Aperture photometry was performed with a variable aperture size that reflected the seeing (1.8$\times$ the full-width at half-maximum of the point spread function), invoking the use of Naylor's optimal photometry algorithm \citep{OptimalPhotometry1998Naylor} when it resulted in an improved overall signal-to-noise ratio for an observing run. The comparison star Gaia DR3 2023676031684657408 ($G=14.7$\,mag) was used. When this comparison star saturated in any instrument for any filter, an alternate comparison star (Gaia DR3 2023676027372561664, $G=18.2$\,mag) was used instead. Both of these stars are non-variable and of similar colour to V407~Vul.

Furthermore, we obtained archival HST Wide Field Camera 3 (WFC3) photometry across the far UV, UV and visible wavelengths. Given that V407~Vul is dominated by a bright G-star in the visible \citep[][]{Steeghs2006V407VulGeminiSpectrum}, the purpose of these observations was to isolate V407~Vul by only capturing a DWD binary that would dominate in the UV. A world-coordinate-system is included with the pointing of the telescope, however, we deem the pointing to be unreliable because of a lack of coordinate solution consistency between frames. In the UV images we are only able to see a single source (the hot WD) and there are no nearby stars to refine the astrometric solution. In the optical we only see a single source (the main sequence star), meaning that we are not able to confirm or deny a direct detection of a spatially separated triple star system from these data alone. However, with the wide spectral coverage covering the UV to near infra-red, we were able to quantify a photometric spectral energy distribution (SED) and isolate that of the binary (Section~\ref{sec:SEDfittingSpectroscopicPhotometric}).


\subsection{Spectroscopy}
The spectroscopy obtained by \citet{Steeghs2006V407VulGeminiSpectrum} clearly shows a prevalent G-type star in the spectrum of V407~Vul. However, we wanted to obtain new spectroscopy for two primary reasons. The first is to have a flux-calibrated spectrum of V407~Vul for accurate fitting of the main sequence star's atmospheric parameters. The second reason is that, after we noticed a UV detection in the HST/WFC3 photometry, we also wanted to obtain spectral coverage blueward of 4100\AA~since here the relative flux contribution of the blue component sharply increases. Therefore, any near-UV excess could be fit. This would put strict atmospheric constraints on the hot WD (Section~\ref{sec:SEDfittingSpectroscopicPhotometric}) and allow a search for consistency on whether the G-type star and the hot WD are interlocked in orbit.

Data were obtained from the Large Binocular Telescope (LBT) on the night 16 June 2025 using the Multi-Object Double Spectrograph \citep[MODS,][]{Pogge2010}. A dichroic splits the beam to be dispersed by a 400 lines/mm grating in the blue and a 250 lines/mm grating in the red. We utilised a 0.6\arcsec\ slit width which gave a resolving power ($R=\lambda/\Delta\lambda$) of $R=1850$ and $R=2300$, respectively. The full wavelength range in this setup was approximately $3400$--$10\,000$\AA~and the blue/red dichroic crossover point occurs at 5650\AA. The data reduction was carried out using the \textsc{modsCCDRed} Python package \citep{Pogge2019} with daytime calibration images. Subsequent processing steps, including wavelength calibration, flux calibration and spectral extraction, were performed with the \textsc{modsidl} pipeline \citep{Croxall2019}.

A spectrum was also obtained using the Keck Observatory using the Low-Resolution Imaging Spectrograph (LRIS) on the night of 22 June 2025. A dual arm setup using the 600/4000 grism in the blue arm and the 1200/7500 grating in the red arm together with a 1.0\arcsec\ slit width gave resolving powers of $R=1100$ and $R=2600$, respectively. This configuration produced a full wavelength range of approximately 3150--7100\AA. 
From prior experience with the setup and reduction, data above 3350\AA~was deemed to be suitable for an accurate instrument response curve correction. Since this data is of very low signal-to-noise anyway, we ignore data below 3350\AA. All LRIS data were wavelength calibrated with internal lamps, flat-fielded, and cleaned for cosmic rays using \texttt{lpipe}, a pipeline for LRIS optimized for long-slit spectroscopy \citep{2019perley_lpipe}.

\section{Timing Solutions}
\label{sec:TimingSolutions}
\subsection{Fitting the photometry}
To quantify the orbital decay of V407~Vul, we adopt the same methodology as outlined in \citet{MundayHMCnc2023}. We started by taking the extracted photometric data from each night and fit Fourier series solutions of the form
\begin{equation}
d + A \cos\left(\omega (t-t_0)\right) + B \sin\left(2\omega (t-t_0)\right) + C \cos\left(2\omega (t-t_0)\right),
\label{eqn:WaveFit}
\end{equation}
where $\omega = 2\pi/P$ is the orbital angular frequency with $P$ the orbital period, $t$ is the centre of exposure time of an individual observation, $t_0$ is the phasing solution to the data, $d$ is a normalised flux offset, $A$ is the amplitude of the fundamental harmonic and $B$ and $C$ are the amplitudes of the first harmonic. A second harmonic is required to compensate for the increase of flux at X-ray peak, which occurs around 0.2~cycles after the peak optical flux. When fitting to each night, $d$, $A$, $B$, $C$ and $t_0$ were allowed to vary. We initially held $\omega$ fixed using the ephemeris of \citet{Barros2007HMCnc}, before using our own, updated ephemeris (Section~\ref{subsec:orbitalEphemeris}) to iteratively refit the photometry. Uncertainties on $t_0$ were deduced by taking the standard deviation of 1000 bootstrapping iterations \citep{Pelisoli2021FastSpinningWD,MundayHMCnc2023}.

We then searched for a trend between timing solutions of unique filters as a function of wavelength. We notice a strong correlation across the SDSS prime and super SDSS datasets, as shown in Appendix~\ref{Appendix:WavelengthDependenceTiming}. A near identical trend is observed in both, where the g$^\prime$/g$_s$ bands are used as a reference point. With that in mind, we decided to combine the timing solutions of all filters of a single night but correcting for this strong colour dependence (including those from non-ULTRACAM/HiPERCAM observations). This was performed by taking the colour-dependent trend from the $g^\prime/g_s$-band (Fig.~\ref{fig:wavelengthDependence}) and solving for the cycle offset with the effective wavelength of each filter. A variance weighted mean was then calculated when V407~Vul was observed with more than one filter simultaneously, and the uncertainty in the offset fit was propagated.

\subsection{Constraining the orbital ephemeris}
\label{subsec:orbitalEphemeris}
\begin{figure*}
    \centering
    \includegraphics[width=\columnwidth, keepaspectratio, clip, trim={0.2cm 0.25cm 0.25cm 0.25cm}]{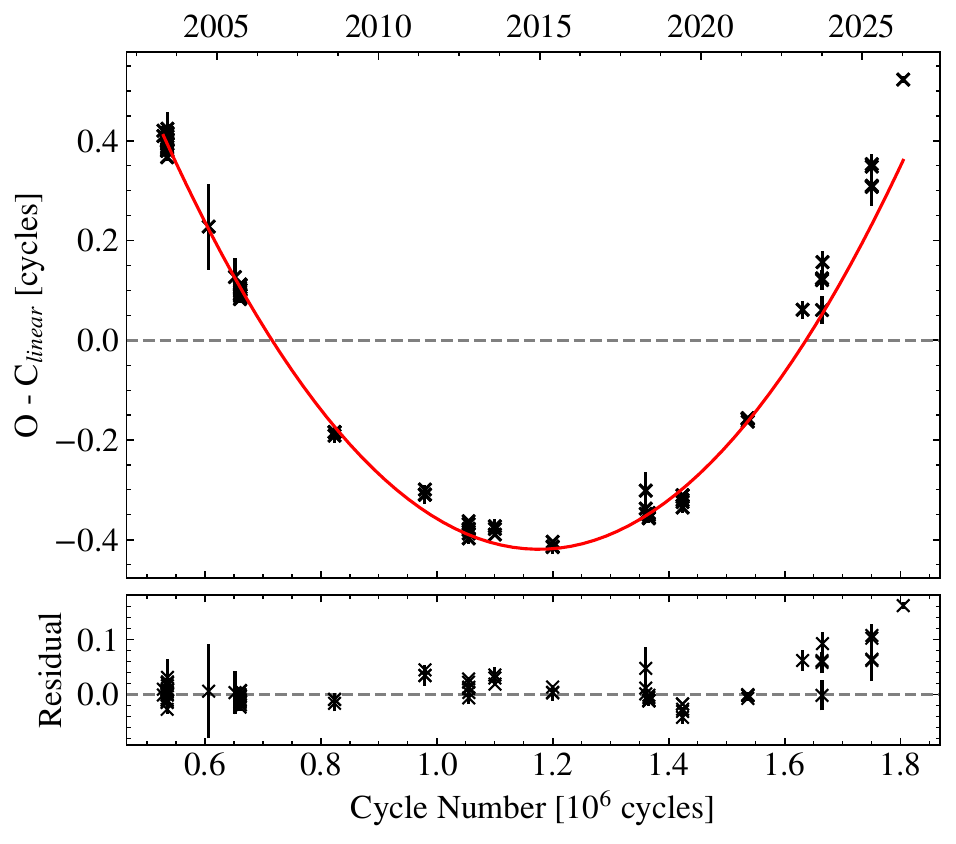}
    \includegraphics[keepaspectratio, clip, trim={0cm 0.15cm 0.25cm 0.25cm}, width=\columnwidth]{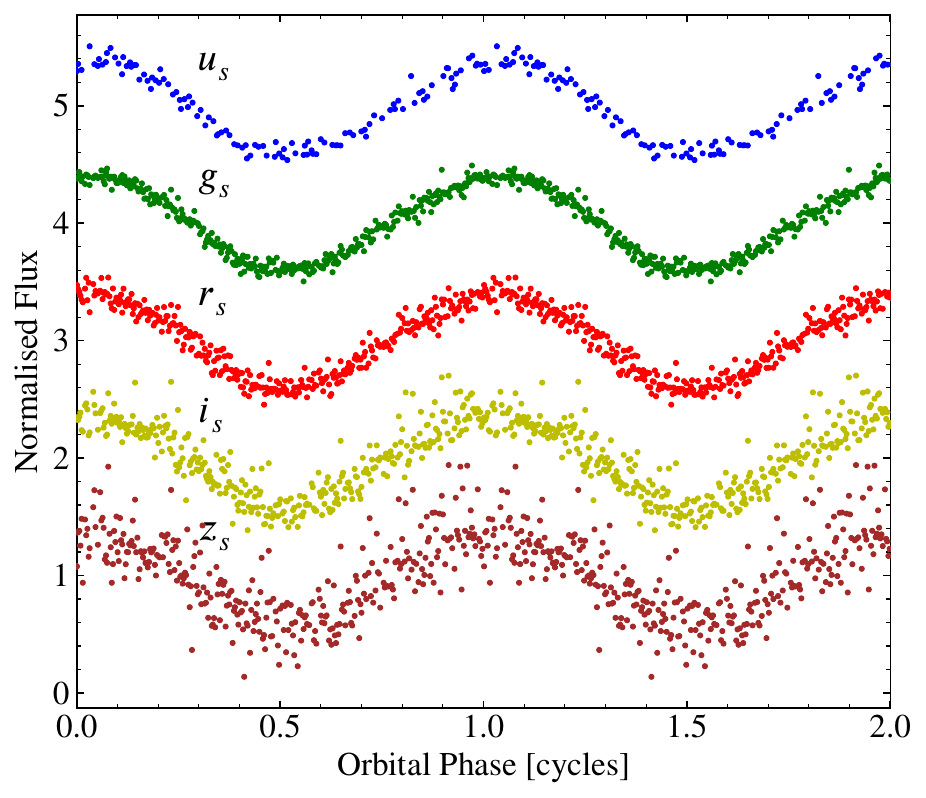}
    \caption{\textit{Left:} Timing residuals to a linear ephemeris for the full set of optical measurements. Both the cycle number and Gregorian year are labelled on the x-axes. In red, a parabola displaying a quadratic fit to the data. For aesthetic purposes, the $O-C_\textrm{linear}$ residuals were computed from an epoch of 49257.533373137 BJD TDB, offset $\phi_0=-2.228$ cycles, frequency $f_0=0.0017562538755$\,Hz. \textit{Right:} The phase folded and normalised light curve of V407~Vul from a single night of HiPERCAM on the GTC in 2021.}
    \label{fig:timings_hcamLCs}
\end{figure*}

The variance weighted mean timing measurements were converted to a cycle number following the epoch of 49257.533373137 to facilitate a direct comparison to other works\footnote{We acknowledge that an epoch near the centre of the full observing baseline would reduce correlation between fitted parameters \citep{MundayHMCnc2023}. The fitted $\dot{f_0}$ is near identical when moving the epoch, and the same series of points above the best fit curve (Fig.~\ref{fig:timings_hcamLCs}) appear for the most recent measurements with a changed epoch too.}. To do so, we used the ephemeris given in \citet{Barros2007HMCnc} to derive the integer cycle under their solution. In no case was a skip in cycles witnessed, which would have been easily noticeable. We then modelled the set of cycle numbers by fitting
\begin{equation}
    \phi\left(t\right) = \phi_0 + f_0 \left(t-T_0\right) + \frac{\dot f_0}{2}\left(t-T_0\right)^2
\label{eqn:PhasingSol}
\end{equation}
with $\phi(t)$ the cycle number, $\phi_0$ a cycle offset from epoch, $f_0$ the frequency at epoch and $\dot f_0$ its first derivative. For a linear (constant frequency) fit, the first two terms on the right of equation~\ref{eqn:PhasingSol} apply, while all terms apply for a quadratic fit. We performed the fit to our data using the \textsc{scipy leastsq} routine, returning uncertainties to a $1 \sigma$ level. The models strongly underfit the data, likely due to some inherent flickering of the mass transferring source \citep{deMiguel2018esCeti, MundayHMCnc2023, Chakraborty2024}, so an extra error of 0.0228~cycles was added to all timing measurements in quadrature to make the reduced $\chi^2$ equal to one. We plot in Fig.~\ref{fig:timings_hcamLCs} the residuals of a linear fit, demonstrating the needed quadratic term\footnote{Inclusion of the Shklovskii effect \citep[][]{Shklovskii1970effect} and galactic rotation to the orbital ephemeris are negligible (see \citet{MundayHMCnc2023} for a discussion).}.

There is a hint of a deviation to a constant period derivative in the most recent timing measurements, lying far off the parabolic fit in Fig.~\ref{fig:timings_hcamLCs}. Based on the scatter of measurements near the centre of the timing dataset (years 2011--2014) that is likely due to underlying stochastic variability of the source, we refrain from investigating the presence of any second derivative term any further. However, we note that this could be an early indication of a second frequency derivative, $\ddot{f}$, which would be rapidly accelerating in magnitude on approach to a period minimum during the AM~CVn turn-around phase \citep{MundayHMCnc2023}. Continued timing measurements of V407~Vul are strongly encouraged to test this hypothesis and probe a $\ddot{f}$ detection, in the optical and in X-rays \citep{Strohmayer2002pdotV407VulXray, Strohmayer2004pdotXray}.

The measured orbital decay ($\dot{f_0}=(1.291\pm0.015)\times10^{-17}$\,Hz\,s$^{-1}$) is now precise to a 1\% level, which in the general-relativistic-only case permits tight constraints on the masses of each star via the system chirp mass. When gravitational wave radiation is the only contribution to the loss of orbital angular momentum, the orbital decay follows
\begin{equation}
    \centering
    \dot f_{GW} = \frac{96}{5}~\pi^{8/3}~\left( \frac{G\mathcal{M}}{c^3}\right)^{5/3}~f_{GW}^{11/3}
    \label{eqn:fdotGR}
\end{equation}
\noindent with $f_{GW}=2f$ the frequency of the gravitational waves, $G$ the gravitational constant, $c$ the speed of light and $\mathcal{M}~=~(m_1\,m_2)^{3/5}~/~(m_1+m_2)^{1/5}$ the chirp mass. For our measured $f_0$ and $\dot f_0$, we obtain an observed $\mathcal{M}=0.1545\pm0.0106\,$\(\textup{M}_\odot\). Mass transfer in V407~Vul acts to oppose inspiral, meaning that this calculation gives a chirp mass minimum only. The true chirp mass of V407~Vul could be much larger since the system lies far from the general relativity prediction of $\dot{P}\propto P^{-5/3}$ for $\mathcal{M}=0.32$\,M$_\odot$ -- a chirp mass that has been witnessed for the majority of the inspiraling, sub-15\,min period, detached/low-mass-transfer-rate DWDs \citep{Chakraborty2024}.

Direct impact accretion and the location of the impact spot itself reduces the allowed accretor-donor mass combinations, so we tested the compatibility of individual star masses and accretion rates with our observed chirp mass. There was no improvement in confining star masses over the combinations permitted by the accretion stream trajectory, meaning that the best information of the two star masses remains in the limits depicted in figure~8 of \citet[][]{Barros2007HMCnc}. The mass limits are therefore in the range of 0.4--0.75\,M$_\odot$ for the accretor WD and 0.1--0.45\,M$_\odot$ for the donor WD, but for a select strip of mass combinations.

\begin{table*}
\centering
\caption{Our best-fit ephemerides for a quadratic and linear ephemeris, solved with equation~\ref{eqn:PhasingSol}. All errors are quoted at a 1$\sigma$ uncertainty. The epoch used in these fits was 49257.533373137, BJD~TDB. }
\label{tab:HMCnc_solution}
\centering
\begin{tabular}{l c c}
\toprule\toprule
Parameter & Quadratic & Linear \\
\midrule
Phase Offset, $\phi_0$ (Cycles) & $0.174\pm0.026$ & $-1.8277\pm0.0072$ \\
Orbital Frequency, $f_0$ (Hz) &  $0.001756245337\pm95$\,pHz  & $0.001756253151\pm12$\,pHz\\
Frequency Derivative, $\dot f_0$ (Hz$\,$s$^{-1}$)& $(1.291\pm0.015)\times10^{-17}$ & -- \\
\bottomrule
\label{tab:ephemerides}
\end{tabular}
\end{table*}

\section{Astrometry}
\label{sec:Astrometry}
A troublesome aspect of V407~Vul has been the presence of a bright G-type star that dominates the visible spectrum \citep[][]{Steeghs2006V407VulGeminiSpectrum}. The natural explanations for this are that the G-type star is in the foreground (although the probability of a chance alignment is strikingly low, see also Section~\ref{subsec:ResultsTripleOrChanceAlignment}), or, the G-type star may be associated to the binary. If associated, this would make V407~Vul a hierarchical triple star, since otherwise a G-type star would engulf a companion at a sub-10\,min orbital period. \citet{Barros2007HMCnc} provide a tentative claim that V407~Vul is a triple system by measuring the pixel variation of a centroid on the detector. They find evidence of positional variation at the same period as the period of V407~Vul, which leads \citet{Barros2007HMCnc} to propose a 0.027\arcsec\ angular separation. If gravitationally bound, the G-type star therefore has a minimum orbital separation of $\approx$30\,AU and the minimum triple orbital period is around 120\,yr.

We revisit an astrometric variability analysis and take advantage of our longest duration of HiPERCAM observations from the night of 16 June 2026, which have a pixel scale of 0.162\arcsec\ (2$\times$2 binning); one half of that of ULTRACAM. The local field around V407~Vul is crowded but the smaller pixel scale permitted us to mask the flux contribution of two nearby contaminant stars almost entirely, whereas for the pixel scale of ULTRACAM this is much more challenging. Furthermore, the observing conditions with this HiPERCAM data had a near-constant image seeing of approximately 0.6\arcsec. There is thus little impact on the amplitude of variability depending on the full width at half maximum of the aperture centroid due to the neighbouring stars in the field with HiPERCAM observations, as was a concern and addressed in \citet[][]{Barros2007HMCnc}.

\begin{figure*}
    \centering
    \includegraphics[clip, trim={0.25cm 0.25cm 0.25cm 0.25cm}, width=\columnwidth]{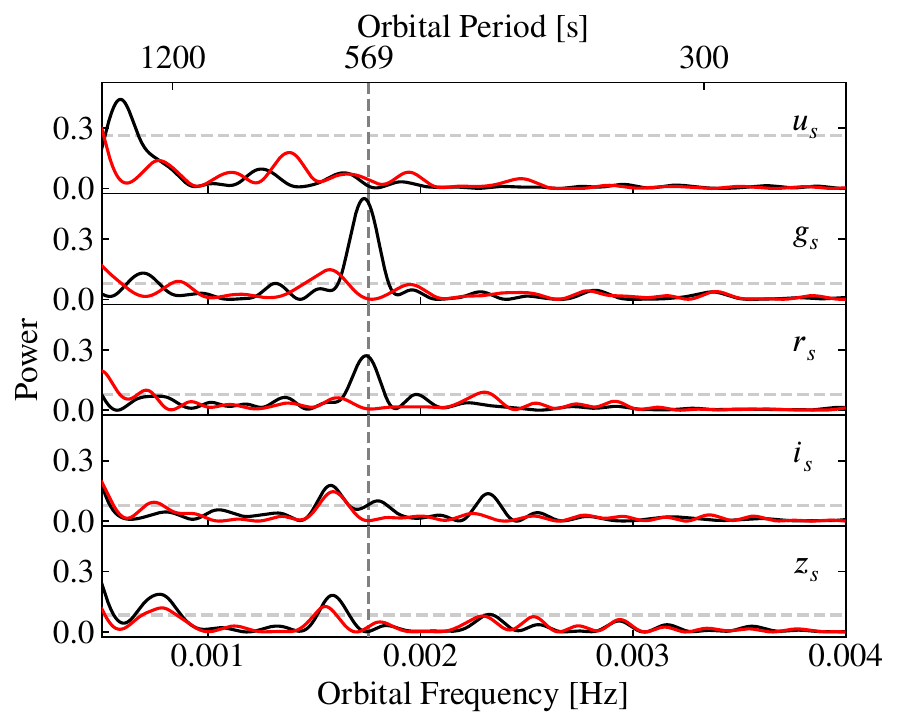}
    \includegraphics[width=\columnwidth, clip, trim={0.25cm 0.21cm 0.25cm 0.25cm}]{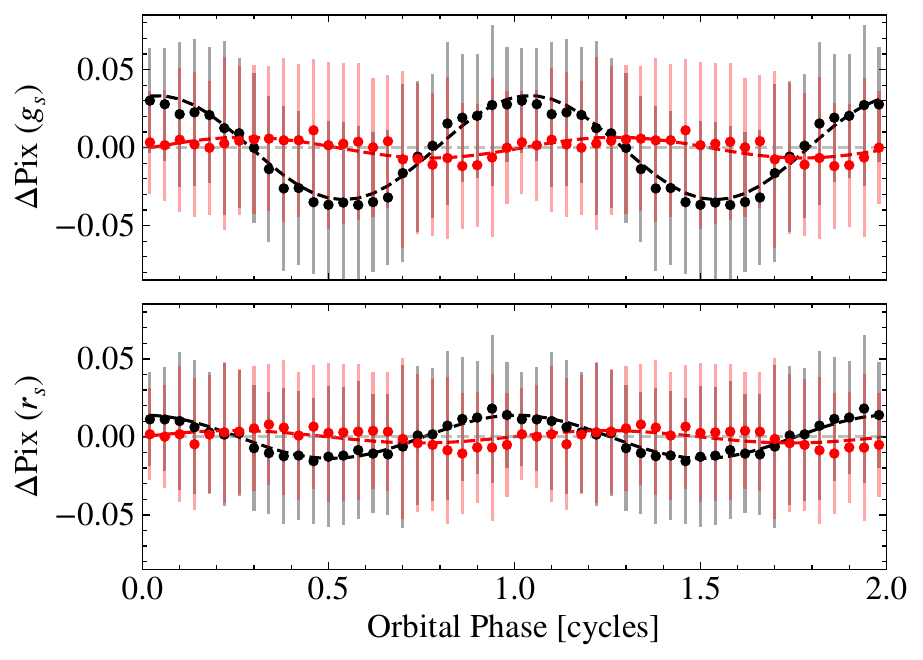}
    \caption{\textit{Left:} Lomb-Scargle periodograms in each HiPERCAM filter, from $u_s$ (top) to $z_s$ (bottom). In red, the periodogram for the Y~pixel deviation and, in black, that for the X~direction. The vertical dashed grey lines are the orbital frequency of V407~Vul, which corresponds to a clear power spike in the X~direction for the $g_s$/$r_s$ bands, and in horizontal dashed grey are 5$\sigma$ false alarm probabilities. 
    \textit{Right:} Folded and binned astrometric variability on a frequency $f=f_0$ for the 16 June 2021 HiPERCAM night. The $g_s$ and $r_s$ pixel deviations are shown at the top and bottom, respectively, with the same colouring system as the left panel. The data has been binned into 25 points with even spacing for clarity.}
    \label{fig:Periodograms}
\end{figure*}

We extracted the pixel location of each aperture in V407~Vul and multiple comparison stars using a symmetric Moffat profile for the super SDSS $u_s$, $g_s$, $r_s$, $i_s$ and $z_s$ data. We then computed the difference in X and Y pixel values between the bright comparisons and V407~Vul. The positional error of the comparison aperture centres (0.15\,pix in $u_s$ and 0.02--0.03\,pix in all other filters per exposure) was negligible compared to that of the V407~Vul aperture (0.15--0.20\,pix per exposure), with photon count being the limiting factor. The X and Y positions showed a trend across the full observation span which we accredit due to differential atmospheric refraction. The airmass of the observations was roughly constant, so we linearly fit this trend to account for a constant times the difference in colour and subtract the trend from the X and Y positions. After, we searched for sinusoidal variability in the pixel location with a single-term Lomb-Scargle periodogram, expecting the astrometric solution to vary on the same period as V407~Vul if a binary star system is spatially separated from the G-type star. These periodograms are plotted in Fig.~\ref{fig:Periodograms}.

We witness a clear detection in the $g_s$ and $r_s$ bands, with no detection in all others because of their lower signal-to-noise ratio data. Modelling the X and Y coordinates with sinusoidal curves offset in phase by 90$^{\circ}$, we obtain pixel variation amplitudes on the orbital frequency of V407~Vul of $0.0340\pm0.0018$\,pix and $0.0144\pm0.0014$\,pix in the $g_s$ and $r_s$ data, while the relative flux amplitude in the light curves were $9.42\pm0.07\%$ and $4.75\pm0.06\%$, respectively. As expected, about twice the relative flux contribution in the $g_s$-band caused twice the pixel amplitude.

For source distances on the magnitude of kpc, these detections clearly indicate that there is a spatial separation between the pulse emitting source and the main sequence star in V407~Vul. 
This discovery thus removes all confusion that the main sequence star is responsible for any of the optical or X-ray variability. Combined with the aforementioned support towards the double degenerate model over the unipolar inductor model, with the nature of the G-type star being the only drawback to the system being settled as an AM~CVn, we consider V407~Vul as having an overwhelmingly large amount of evidence in favour of being a AM~CVn binary. We promote this interpretation for all future mentions of the source\footnote{With an AM~CVn model, the relative contribution of AM~CVn should be strongest in the $u_s$-band, however the lack of astrometric wobble detected here is unsurprising because of the lower signal-to-noise data. Similarly moving to the redder end of the spectrum, the dilution of the main sequence star makes an astrometric wobble in the $i_s$- and $z_s$-band too small to show any significant sign of an oscillation with ground-based observations}.

At this stage of our investigation with the information presented, we stress that the AM~CVn and the main sequence star could be gravitationally bound or a chance alignment. This however will be resolved in Section~\ref{subsec:ResultsTripleOrChanceAlignment}. \citet{Kupfer2024lisa} remark that V407~Vul shows no indication of any astrometric wobble noise in \textit{Gaia}~DR3, but emphasise that the sensitivity to an orbiting tertiary object under the current \textit{Gaia} time baseline of only a few years is low. The orbit of a pulse emitting binary and the main sequence star would not be detectable by \textit{Gaia} for any source distance on the magnitude of a kpc \citep{ElBadry2024GaiaReview}, limited by the 0.1\arcsec\ diffraction limit of Gaia \citep{GaiaMissionPaper}. Hence, a triple system existing even when there are no astrometric anomalies in \textit{Gaia} would not be unexpected. Perhaps \textit{Gaia}-like successors \citep[e.g.][]{Hobbs2018} will be able to resolve the individual sources of light.


\section{Spectro-photometric fitting}
\label{sec:SEDfittingSpectroscopicPhotometric}
We combine photometry extracted from the HST/WFC3 imaging with survey data obtained from CDS\footnote{\url{https://cds.u-strasbg.fr/}} and flux calibrated ULTRACAM SDSS prime magnitudes from \citet{Barros2007HMCnc} to fit the photometric SED of V407~Vul. A full list of flux measurements incorporated are supplied in Appendix~\ref{appendix:Photometry}. We additionally extracted archival photometry from the Neil Gehrels Swift Observatory to verify the flux calibration of our UV photometry, showing consistency. However, we do not include the Swift Observatory exposures in the SED fitting as the exposure times were not integer values of the orbital period, and a small number of observing epochs means that a mean target brightness would be unreliable.

\subsection{Reddening}
\label{subsec:reddening}
V407~Vul is a highly reddened source, making the fitting of data challenging both now and historically \citep{Haberl1995V407VulDiscovery, Ramsay2002v407VulSpectAndPhot, Ramsay2005v407Vulxray, Ramsay2006v407Vulxray, Steeghs2006V407VulGeminiSpectrum}. Many methods were tested to find the best prescription of reddening in our SED fitting. These were: as a free parameter, fixed to tabulated values or a free parameter but weighting the solution towards tabulated values with priors. All were tested for varying E(B$-$V) extinction coefficients with $R_V$ fixed at 3.1 or $R_V$ allowed to be a free parameter. As well, each of these free versus fixed parameter considerations were investigated when abstracting the inner binary with a one-star or two-star model. However, ultimately, the combination of relatively low signal-to-noise data in the UV (due to the $\approx$22--23\,mag faintness in this wavelength range) and the use of an input spectral template that does not include e.g. emission lines (which is fundamentally inaccurate) causes many fitting degeneracies.

A selection of extinction coefficients were trialed through the procedure described in Section~\ref{subsec:fittingTechnique}, which inherently assumes that the main sequence star and the AM~CVn are located at the same distance. No model could provide anywhere near a good fit to the optical spectrum or the HST/WFC3 data for an extinction of $A_V=5.6$\,mag \citep[][E(B$-$V$)\approx1.8$\,mag]{Cropper1998v407vulDiscovery}, with the poor, best-fitting solutions tending towards an accretor in excess of 100\,000\,K and a distance around 1\,kpc. When attempting an E(B$-$V$)=0.50$\,mag, which is a lower limit predicted by \citet{Steeghs2006V407VulGeminiSpectrum}, the fitted parameters are very similar, where the accretor radius pushes to $R_\textrm{WD}\approx0.03$\,R$_\odot$, the main sequence star temperature increases slightly and so does the source distance to approximately 4\,kpc. Increasing the extinction further to E(B$-$V$)\geq0.60$\,mag, the accretor again requires a temperature in excess of 100\,000\,K. Finally, we inspected the consequence of E(B$-$V$)=0.30$\,mag and expectedly notice a large flux excess in the UV. The lack of consistency with new and improved data demonstrates that the extinction of V407~Vul have likely been overestimated in the past decades. An extinction coefficient of E(B$-$V)$=0.4$--0.5\,mag appears most suitable.

We witnessed a large flux deficiency in the blue spectra when fitting with a single main sequence star model. When fitting just the photometry and spectra $<4000$\AA~with all reddening tests above, isolating the majority of flux as being from the AM~CVn, we found distances that are completely consistent with the results of fitting all data together. Fitting all data with different distances for the AM~CVn and the main sequence star, we find consistent solutions also. The sources therefore must be found at a similar distance. This lends extra evidence towards the objects being bound as a triple star system rather than the AM~CVn component being a distant background source, and so we fit the main sequence star and the AM~CVn with a consistent distance.

We determined that the best prescription of reddening would be to interpolate reddening map values, fixing $R_V=3.1$, and that we would instead allow more lenience in the number of free parameters used for stellar parameters (Section~\ref{subsec:fittingTechnique}). We take advantage of recent reddening maps constructed from \textit{Gaia}~DR3 and 2MASS photometry \citep{Lallement2022} to predict the colour excess, E(B$-$V), which extend to 3.6\,kpc in the line of sight of V407~Vul. The extinction changes very little between 2.5\,kpc and 3.6\,kpc, with a linearly increasing trend. So, we fit a straight line to this data to interpolate/extrapolate E(B$-$V) for trial distances. The E(B$-$V) used in our fitting is thus in the range of 0.43--0.44\,mag, consistent with other reddening maps as well \citep[e.g.][]{Green2019PanstarrsReddening}. We recommend caution to be taken so as not to over-interpret the results of the AM~CVn component because of a lack of clarity in the exact extinction coefficient, since these are results rely on the bluer/UV wavelength data. The donor star is largely unconstrained for this reason, while the approximate solution of the accretor can be assumed.

\subsection{Model atmospheres}
\label{subsec:spectrophotometricModelAtmospheres}
We model the bright main sequence star with high-resolution \textsc{PHOENIX} synthetic spectra \citep{Husser2013PHOENIX}. The grids were linearly interpolated for a $\log g = 4.4$\,dex star, which is a typical surface gravity for a late G-type main sequence star. Expecting the binary system to be hydrogen poor as an AM~CVn, we model both stars with a pure He atmosphere with the spectra of \citet[][]{ElenaCukanovaite2021DBmodels} when the effective temperature is less than 40\,000\,K or the DO~WD models of \citet[][]{Bedard2020MontrealWDModelsMTR} for higher trial temperatures. We fix the spectral shape to that of a $\log\text{(g)}=8.0$\,dex WD for both grids to remove fitting degeneracies, with $\log(\textrm{g})$ having a far smaller impact on the integrated flux over a bandpass than the source temperature. We also omit metal lines, which frequently appear in hot WDs \citep{Filiz2026}.  Of course, neither WD grid perfectly replicates the true AM-CVn like spectrum of V407~Vul and ignores all emission lines that may be present in the optical spectrum, yet are washed-out by the main sequence star. Similarly, any potential emission lines in the UV are not considered in the input template spectrum when fitting the HST photometric data points.

\subsection{Fitting technique}
\label{subsec:fittingTechnique}
The flux-calibrated spectra and photometry from Pan-STARRS \citep{Panstarrs}, SDSS \citep{SDSSdr16}, the XMM-Newton optical/UV monitor \citep{XmmUV2001}, HST and \citet{Barros2007HMCnc} were simultaneously fit with the Keck/LRIS and LBT/MODS spectra. However, the Keck/LRIS and LBT/MODS spectra were handled separately to isolate potential inaccuracies when correcting the instrument response function in the near-UV, generating two unique sets of fitted parameters. The utilised photometry was identical in each case. Furthermore, we wanted to maintain colour information in the flux-calibrated spectra while not suffering from absolute flux inaccuracies that are especially relevant for ground-based observations, such as slit losses. So, we fit the overall shape of the spectra but arbitrarily scaled the synthetic fluxes through a $\chi^2$ minimisation. This is not a concern for the photometry, hence physical trial distances were used to scale the absolute photometry to an observed flux. Cross-correlation of the normalised Keck/LRIS and LBT/MODS spectrum with normalised model templates indicated a radial velocity of $-132$\,km\,s$^{-1}$ in both cases, which was used to offset the model to the data. Certain spectral signatures were masked that are contaminated by the interstellar medium, being the Na~I~D doublet (5885--5900\AA) and diffuse interstellar bands at 5780\AA, 5797A\AA~and 6283\AA~also.

We employed a Markov Chain Monte Carlo (MCMC) algorithm using the python package \textsc{emcee} \citep[][]{emcee2013}. The radius of the main sequence star was guided by interpolating the temperature-radius values of spectral types in Table~7 of \cite{eker2018MRRMTR} and setting a Gaussian prior with standard deviation 0.02\,R$_\odot$. This standard deviation was chosen as it well reflects the scatter between adjacent spectral types for the \citet{eker2018MRRMTR} values. The model grids were scaled from an Eddington flux to a flux at Earth with the trial reddening constant applied, before integrating over the filter transmission profile. 100 walkers with 1500 burn-in and 500 post-burn-in iterations were used, and the post-burn-in results of the MCMC analysed.

Initially, the free parameters in the MCMC for the main sequence star were temperature ($T_\textrm{MS}$), radius ($R_\textrm{MS}$), [Fe/H] and [$\alpha/$Fe] the element abundance. The free parameters for each WD were the temperature ($T_\textrm{WD1}, T_\textrm{WD2}$) and radius ($R_\textrm{WD1}, R_\textrm{WD2}$), with the accretor being star 1 and the donor being star 2. The distance to the system ($D$) was varied freely as well. When doing this, we quickly realised that the donor WD contributes about 1\% of the total flux only. Hence, the stellar parameters struggled to converge, and the only inference that we could obtain was that the donor has a maximum effective temperature of 15\,000\,K. While the final posterior distributions were wide and $R_\textrm{WD2}$ was largely unconstrained, the best-fit values to both datasets were approximately $T_\textrm{WD2}=8000\,\textrm{K}$ and $R_\textrm{WD2}=0.035\,\textrm{R}_\odot$. In order to decrease the number of degrees of freedom while still wanting to incorporate the smaller amount of flux from the donor for a more physically accurate methodology, we decided to take and fix these values. The final MCMC parameters that we fitted were therefore $T_\textrm{MS}$, $R_\textrm{MS}$, [Fe/H], [$\alpha$/Fe], $T_\textrm{WD1}$, $R_\textrm{WD1}$ and $D$.

\section{Results and discussion}
\label{sec:ResultsAndDiscussion}
\subsection{Spectro-photometric fit parameters}
\label{subsec:SEDfitparameters}
The Keck/LRIS and LBT/MODS datasets were handled separately and hybridly fit with the photometric data to isolate any reduction and/or flux calibration errors. All fitted parameters are listed in Table~\ref{tab:fittedParameters}, with the fits overlaid on the data in Fig.~\ref{fig:spectraPhotometryFits}. Corner plot diagrams showing the covariance between independent variables and the resultant parameter posterior distributions are plotted in Appendix~\ref{appendix:Corners}. 

The two fits produce very similar atmospheric parameters. The flux excess from the AM~CVn component is clearly noticeable in both spectra, and we are able to obtain solutions that well fit the spectra and photometry simultaneously. The best spectral fit is found for the LBT/MODS data by obtaining a cooler main sequence star temperature and larger WD radius than in the Keck/LRIS solution. The WD is the dominant flux contributor below $4000$\AA, but we see that the photometry becomes worse fit for the LBT/MODS case. A still very good spectral fit is obtained for the Keck/LRIS solution while being able to model the HST photometric points better. Moreover, if we plot the Keck/LRIS solution over the LBT/MODS data and vice versa, the shape of the spectral energy distribution is well suited.

The fact that we are able to obtain very similar solutions using data from two unique instruments, sites and reduction pipelines gives us confidence in the near-UV spectral data quality. Because of the various assumptions and abstractions that go into the spectro-photometric fitting, it is best to combine the results instead of favouring one solution. We do this by concatenating the distributions obtained for each parameter then taking its median and 68\% confidence interval for the median and 1$\sigma$ error \citep[e.g.][]{Munday2025}. These values are also presented in Table~\ref{tab:fittedParameters}, and all further discussion of the results refers to these `adopted' parameters.

{\renewcommand{\arraystretch}{1.25}
\begin{table}
    \caption{Stellar parameters for the main sequence star and the accretor WD in the AM CVn. Spectra were simultaneously fit with absolute flux measurements (including our new HST data points). Formal errors are given for the Keck/LRIS and LBT/LRIS solutions and the true uncertainties would be much larger. The adopted values come from summing the posterior probability density functions from each fit and then sampling the resulting distribution 5\,000 times to construct an `adopted' posterior distribution. Errors are quoted at a 1$\sigma$ significance. The fitted results come after including the flux contribution of a $T_\textrm{WD2}=8000$\,K and $R_\textrm{WD2}=0.035$\,R$_\odot$ donor WD.}
    \centering
    \begin{tabular}{l l l l}
        \toprule \toprule
         & Keck/LRIS & LBT/LRIS & Adopted\\
         \midrule
$T_\textrm{MS}$ [K] & 5100$^{+30}_{-20}$ & 4940$^{+10}_{-10}$ & 5040$^{+60}_{-110}$  \\ 
$R_\textrm{MS}$ [R$_\odot$] & 0.92$^{+0.02}_{-0.02}$ & 0.92$^{+0.02}_{-0.02}$ & 0.92$^{+0.02}_{-0.02}$  \\ 
$T_\textrm{WD}$ [K] & 59300$^{+6700}_{-6300}$ & 57900$^{+3400}_{-2800}$ & 58100$^{+5500}_{-3900}$  \\ 
$R_\textrm{WD}$ [R$_\odot$] & 0.021$^{+0.001}_{-0.001}$ & 0.024$^{+0.001}_{-0.001}$ & 0.023$^{+0.001}_{-0.002}$  \\ 
$D$ [pc] & 3600$^{+100}_{-100}$ & 3450$^{+80}_{-80}$ & 3510$^{+140}_{-110}$  \\ 
{[Fe/H]} & -0.80$^{+0.08}_{-0.08}$ & -0.79$^{+0.03}_{-0.03}$ & -0.80$^{+0.05}_{-0.05}$  \\ 
{[$\alpha/$Fe]} & 0.31$^{+0.05}_{-0.06}$ & 0.18$^{+0.03}_{-0.03}$ & 0.23$^{+0.11}_{-0.06}$  \\ 
\bottomrule
    \end{tabular}
    \label{tab:fittedParameters}
\end{table}
{\renewcommand{\arraystretch}{1}

The main sequence star was found to have T$_{\text{eff}}=5040^{+60}_{-110}$\,K, is metal-poor with $\textrm{[Fe/H]}=-0.80^{+0.05}_{-0.05}$, and is slightly $\alpha$-enhanced with $\textrm{[Fe/H]}=0.23^{+0.11}_{-0.06}$. We obtain a main sequence star radius of 0.910\,\(\textup{R}_\odot\), which corresponds to a mass of approximately 0.9\,M$_\odot$ \citep[][]{eker2018MRRMTR}{}{}. Hot accretor temperatures are expected in DWD AM~CVns. The temperature is primarily constrained by a combination of the mass of the accretor WD and the mass transfer rate of the accreted helium-rich material. The material settles on the accretor, becomes deeply compressed in the envelope and releases heat in the process. We find that the accretor in V407~Vul has an effective temperature of 58\,100$^{+5500}_{-3900}$\,K, which aligns well with that predicted in theoretical evolutionary sequences of AM~CVn binaries \citep{Bildsten2006, Wong2021AMCVnMESAPureHe}. A comparison of our measured accretor effective temperature with similar evolutionary tracks to those presented in \citet{Wong2021AMCVnMESAPureHe}, for the range of permitted initial donor and accretor masses of V407~Vul, predicts that the present-day mass transfer rate is $\dot{M}=10^{-8}$~--~$10^{-7.5}$\,M$_\odot$. Higher present-day mass transfer rates would require the accretor temperature to have been rapidly increasing (within $\approx10^4$\,yr), so it is unlikely that we spot the system in such a transitional state. Smaller mass transfer rates would be unlikely since the observed temperature would be reached after $\approx10^7$\,yr, which is the approximate AM~CVn evolutionary timescale. Moreover, we note that the effective accretor temperature is significantly lower than the $\gtrapprox10^5$\,K surface temperature needed for the assumed, near-perfect conductivity for the unipolar inductor model \citep{Wu2002UImodel, Wu2009review}.

We are not able to provide much physical insight into the donor or the composition of the accreted material since there is little spectral signature of the AM~CVn binary. This would require UV spectra, whereby the donor structure could be investigated from relative carbon-nitrogen-oxygen abundances in the accreted material \citep{Chakraborty2024} or any other metal lines with improved spectroscopy. Based on our photometric and spectroscopic modelling, $T_\textrm{WD2}>15\,000$\,K would poorly fit the observations, since the extra UV flux becomes excessive while the accretor temperature is still needed to remain at approximately 60\,000\,K to generate enough flux in the spectra at 3500--4000\AA. As noted in Section~\ref{subsec:fittingTechnique}, this means that the donor can be assumed to have a maximum effective temperature of 15\,000\,K. When we let the donor parameters be free variables, a best solution is found for donor parameters $T_\textrm{WD2}=8000$\,K and $R_\textrm{WD2}=0.035$\,R$_\odot$, but for a wide distribution of possible atmospheric parameters.

We obtain a source distance of 3510$^{+140}_{-110}$\,pc. This is on the upper edge of the approximate distance limits of $D=1100$--3500\,pc found in \citep{Steeghs2006V407VulGeminiSpectrum}, albeit that they used a slightly different reddening prescription. Using the reddening that we find, the distance bounds would slightly increase, thus our measurement aligns with the expectation of these authors. Importantly, this distance is much larger than using the approximate distance that has been adopted in recent ``verification binary'' gravitational wave predictions \citep{Kupfer2018LISAverificationBinaries, Finch2023, Kupfer2024lisa}. Updated gravitational wave predictions will be described in Section~\ref{section:GravitationalWaveSource}. As a test, we enforced multiple hotter main sequence star temperatures around 5200\,K and checked for model consistencies with the two WD models but letting the reddening extend freely to higher extinction coefficients to inspect the range of permitted distances, always desiring a higher extinction coefficient and slightly further distance in the range 3800--4200\,pc.

\begin{figure*}
    \centering
    \includegraphics[width=14.5cm, clip, trim={0.6cm 0.5cm 0.5cm 0.5cm}]{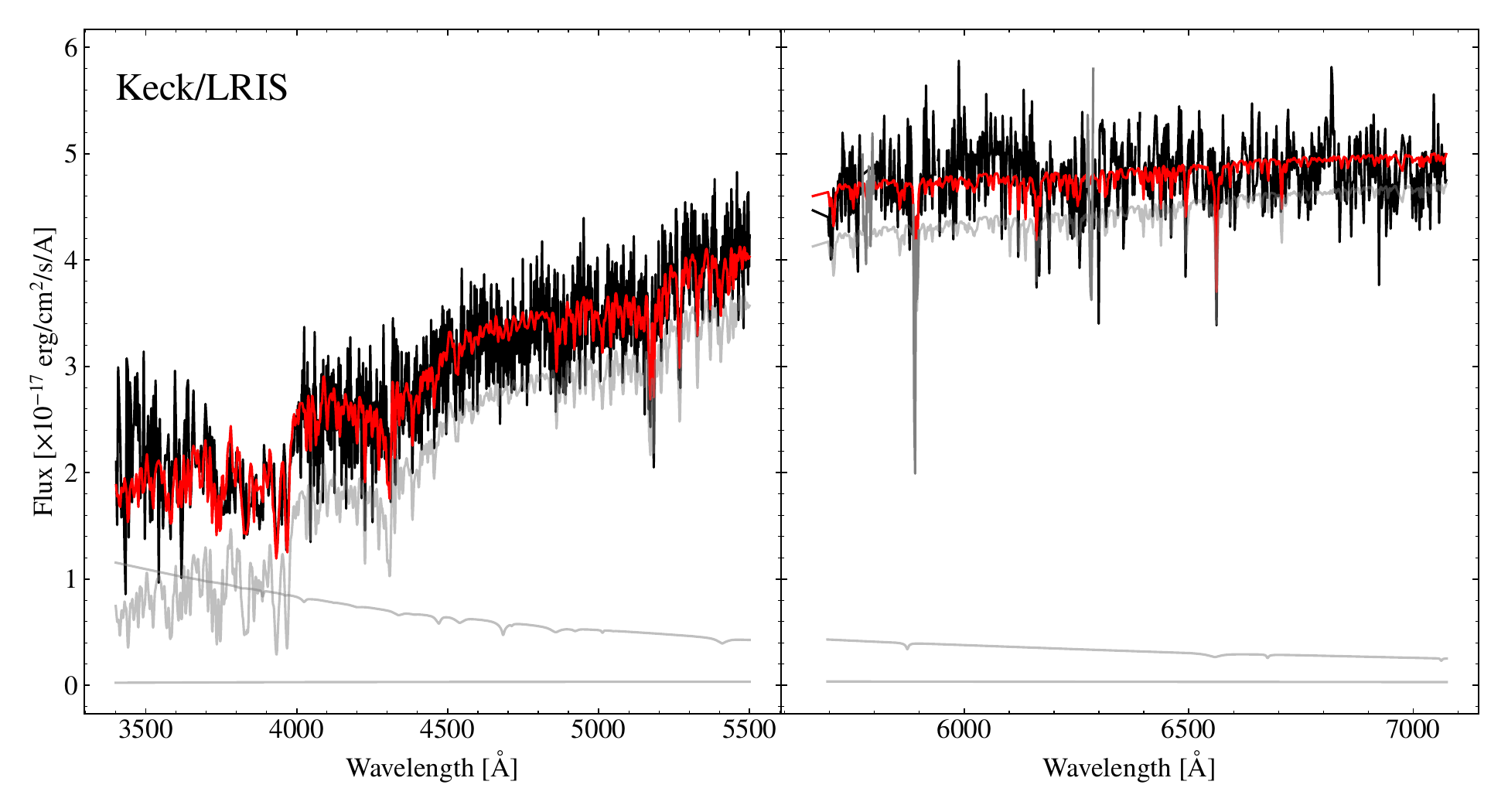}
    \includegraphics[width=14.5cm, clip, trim={0.6cm 0.5cm 0.5cm 0.5cm}]{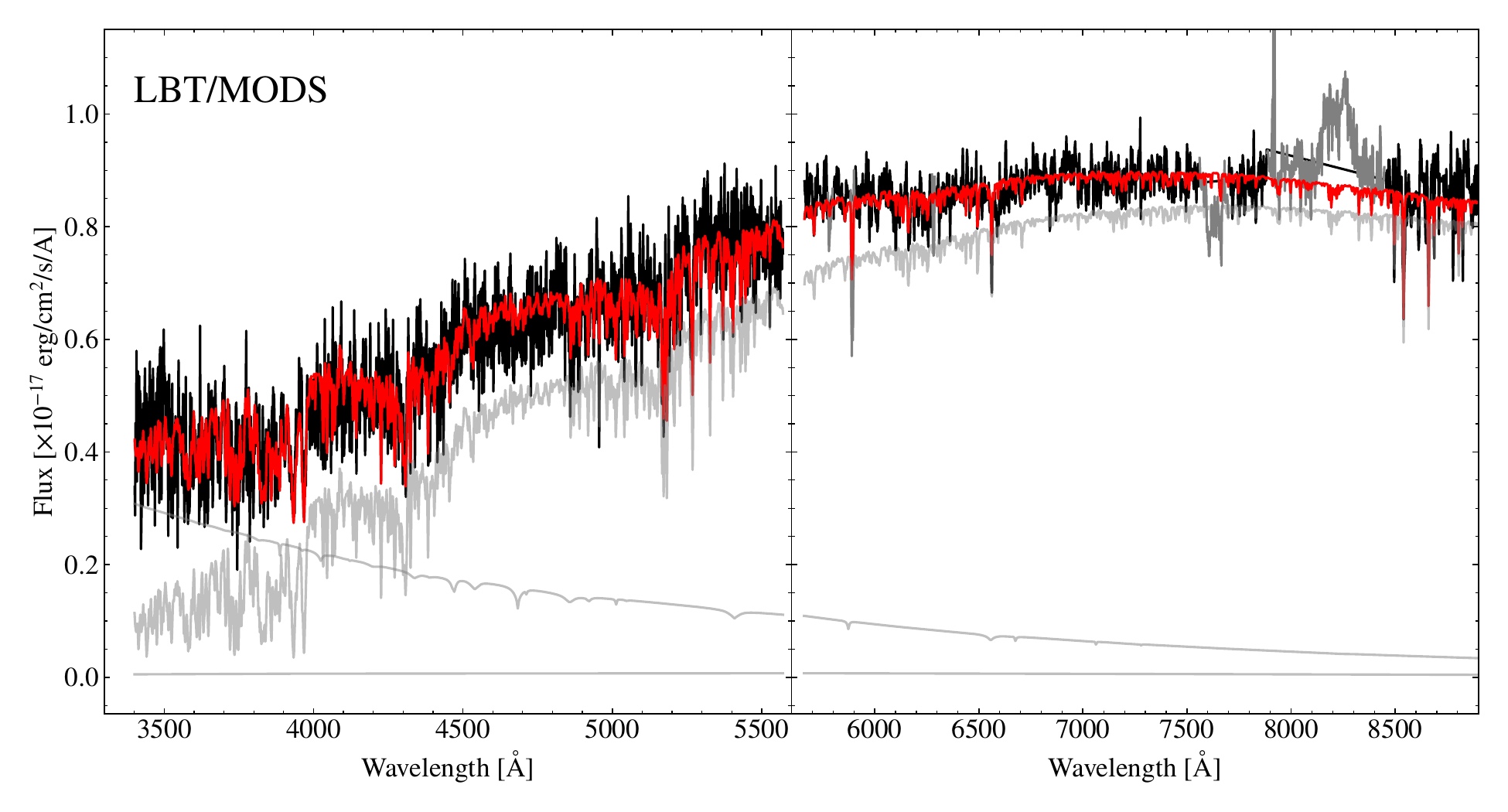}
    \includegraphics[trim={0cm 0.5cm 1.cm 2cm},clip,width=8.07cm]{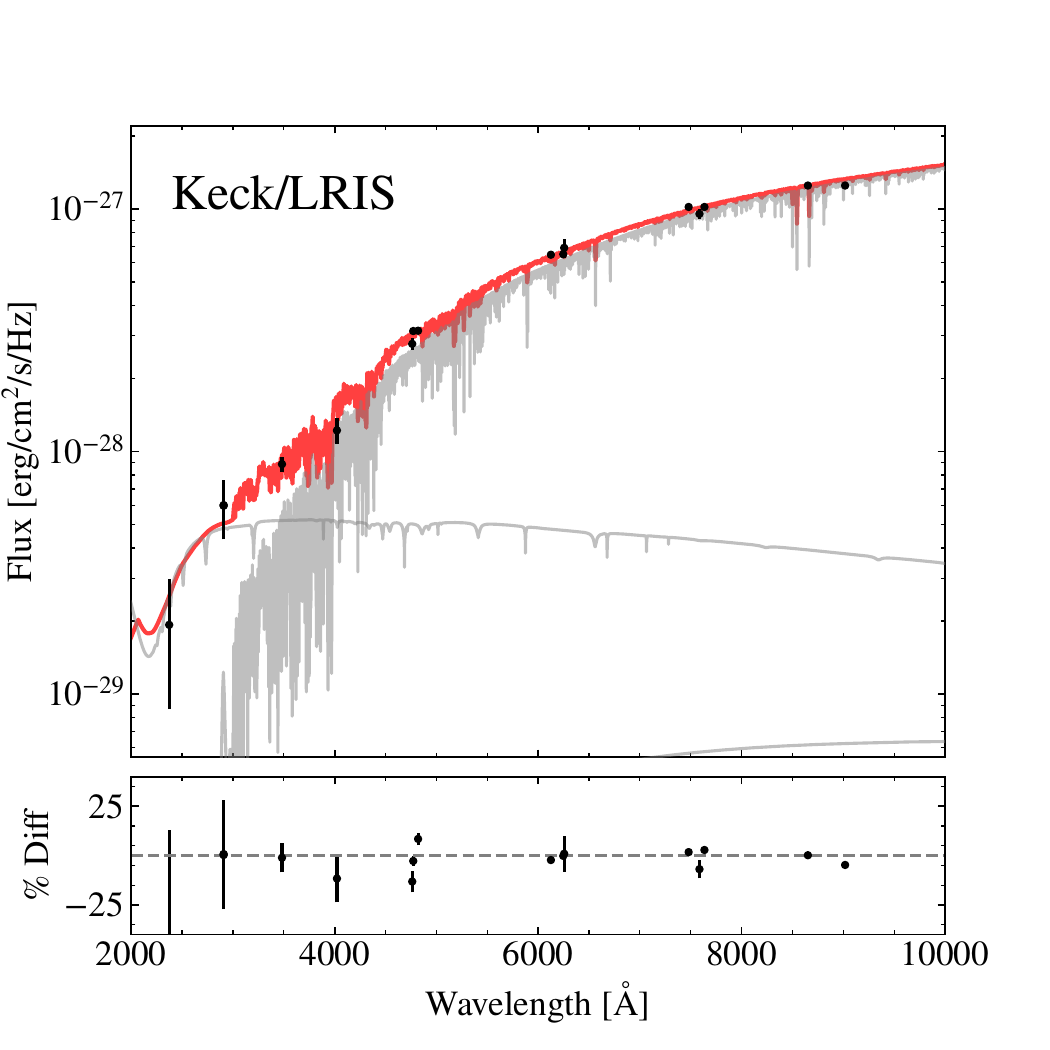}
    \includegraphics[trim={0cm 0.5cm 1.cm 2cm},clip,width=8.07cm]{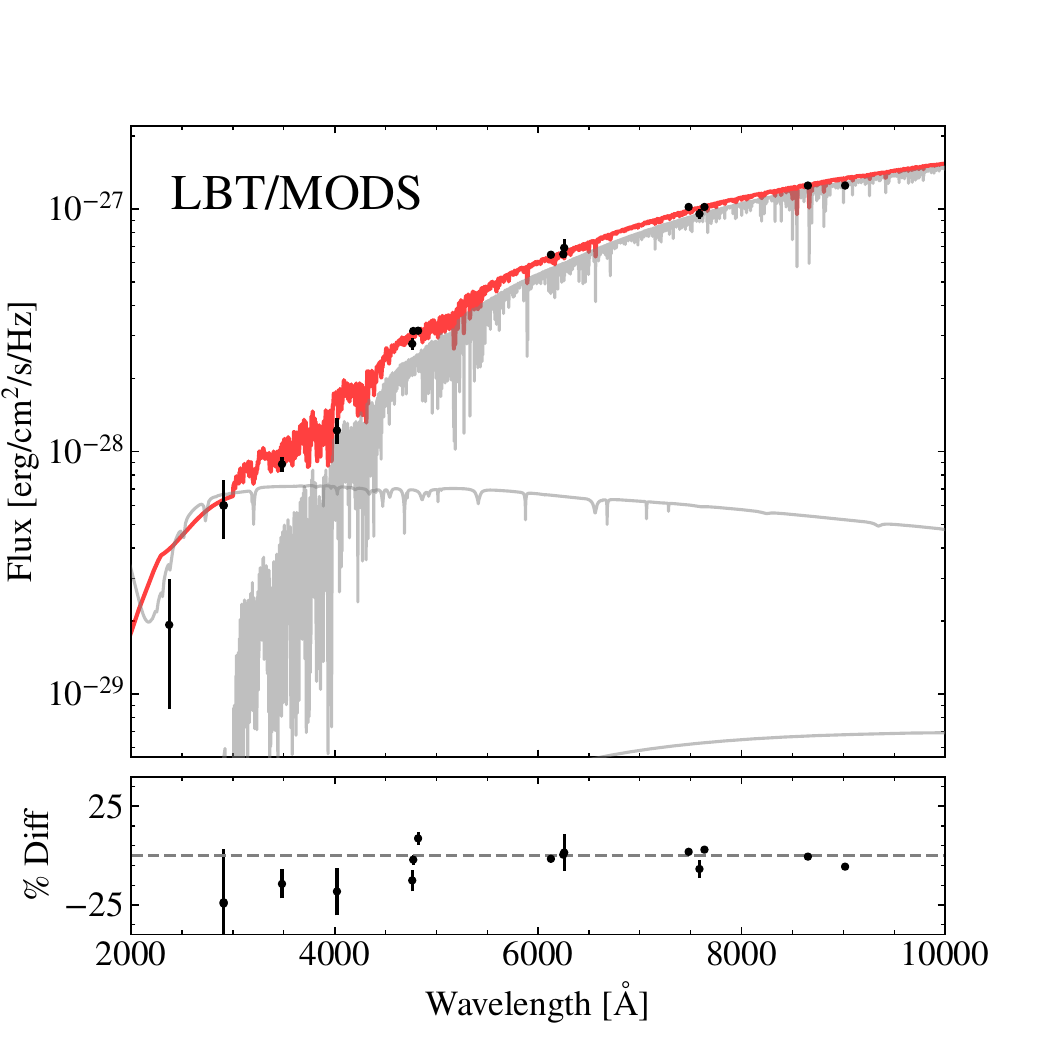}
    \caption{The unique fits to the Keck/LRIS (top) and LBT/MODS (middle) data, split into the blue and red arms. Below, the two solutions on the photometric SED of V407~Vul. The spectra and photometry for each labelled dataset were fit simultaneously. All plots show the observations in black, the combined model in red and the flux contributed from the main sequence star, the accretor and the donor WDs in grey. The LBT/MODS spectrum has greyed regions that are frequent pipeline artefacts, masked completely. The combined model in the SED plots has been smoothened for clarity. The percentage difference between the integrated and observed flux in a filter is plotted beneath the SEDs. $T_\textrm{WD2}=8000$\,K and $R_\textrm{WD2}=0.035$\,R$_\odot$ is assumed for the donor WD.}
    \label{fig:spectraPhotometryFits}
    
\end{figure*}

\subsection{Triple star system or chance alignment?}
\label{subsec:ResultsTripleOrChanceAlignment}
As the centre of light between the main sequence star and the AM~CVn varies depending on the relative brightness of each, the pixel amplitude in each passband, denoted $A$, is non-constant and does not alone represent the angular separation, $d$, of the binary. It is related to the fractional amplitude of flux variation in that passband, $V$, and the fractional contribution of the main sequence star at minimum flux, $f_\textrm{MS,min}$, via
\begin{equation}
    A = \left(\frac{V}{V+1}\right) f_\textrm{MS,min}~d
    \label{eqn:fractionalFluxAmplitude}
\end{equation}
We integrated the adopted SED model over each HiPERCAM passband to predict the average percentage flux contribution of the AM~CVn and the main sequence star, while $A$ and $V$ are measured properties from the observations. We then used the averaged fluxes and the photometric amplitude to compute $f_\textrm{MS,min}$. Inserting these into equation~\ref{eqn:fractionalFluxAmplitude} solves for a predicted spatial separation. The full set of measurements for this analysis is presented in Table~\ref{tab:periodTriple}. The angular separation between the two sources of light is approximately 0.03--0.04\arcsec, and the mean is $0.0346\pm0.0018$\arcsec.

{\renewcommand{\arraystretch}{1.1}
\begin{table*}
    \caption{The deduced triple system parameters in each super SDSS filter. Results from Sections~\ref{sec:Astrometry} and \ref{sec:SEDfittingSpectroscopicPhotometric} are combined, while the angular separation on sky ($d$) is determined from the percentage flux contribution of the main sequence star at photometric minimum (f$_\textrm{MS,min}$) combined with equation~\ref{eqn:fractionalFluxAmplitude}. f$_\textrm{MS,min}$ was computed through f$_\textrm{MS,avg}$ and the flux amplitude of the light curve ($V$). The spatial separation is calculated using a distance of 3510$^{+140}_{-110}$\,pc.} 
    \centering
    \begin{tabular}{c c c c c c c}
    \toprule\toprule
Filter & $A$ [pix] & $V$ [\%] & f$_\textrm{MS,avg}$ [\%] & f$_\textrm{MS,min}$ [\%] & $d$ [$^{\prime\prime}$] & Sky Separation [AU]\\
\hline
$u_s$ & -- & $21.74\pm0.33$ & 40.23 & 51.40 & -- & --\\
$g_s$ & $0.0340\pm0.0018$ & $9.42\pm0.07$ & 78.96 & 87.17 & $0.0406\pm0.0021$ & $142.4\pm7.4$\\
$r_s$ & $0.0144\pm0.0014$ & $4.75\pm0.06$ & 90.42 & 94.94 & $0.0285\pm0.0028$ & $100.0\pm9.7$\\
$i_s$ & -- & $3.07\pm0.07$ & 94.27 & 97.26 & -- & --\\
$z_s$ & -- & $2.24\pm0.09$ & 95.98 & 98.18 & -- & --

    \end{tabular}
    \label{tab:periodTriple}
\end{table*}
{\renewcommand{\arraystretch}{1}

While resolving individual sources of light in the triple system is not feasible for the foreseeable future (Section~\ref{sec:Astrometry}), we can use the ground-based astrometric results to search for compatibility of a stationary background source being the site of photometric variability. \citet{Barros2007HMCnc} found $f_{\rm MS,min}d=0.0192\pm0.0046$ in the $g\prime$-band based on ULTRACAM observations in August 2005. Integrating our adopted model over the SDSS~g$^\prime$ filter, the percentage flux of the main sequence star at photometric minimum is 87.9\%. The revisited \citet{Barros2007HMCnc} angular separation becomes $d=0.022\pm0.005$\arcsec. The maximum difference between our and their measurements would hence be when the AM~CVn passes directly through the centre of light of the G-type star between these dates (or vice versa), giving $0.056\pm0.006$\arcsec. \textit{Gaia} reports a proper motion of the unresolved triple of $\mu_\alpha=-2.09\pm0.16$\,mas\,yr$^{-1}$ and $\mu_\delta=-4.79\pm0.23$\,mas\,yr$^{-1}$, and our data originates from June 2021. The net angular motion of V407~Vul in this time would be 
$0.0763\pm0.0032$\arcsec. Therefore, comparing the maximum difference between angular separations and the total proper motions, V407~Vul is incompatible with a stationary component being present to at least a $3.2\sigma$ level.

As a second test and especially relevant in a crowded field, there is also the similar case for the individual constituents of V407~Vul where we can test the likelihood that the AM~CVn binary and the main sequence star are not gravitationally bound. Here, the AM~CVn would be a separate background source and the main sequence star would be in the foreground (or vice versa) and the objects must have to move in a similar direction with a similar proper motion for angular separation $d$ consistency. We addressed this situation by simulating stars within 0.1\,deg$^2$ in the direction of V407~Vul using \textsc{TRILEGAL} \citep{TRILEGAL2005}, while setting the total $A_V$ extinction along the line of sight to be what we measure for V407~Vul (Section~\ref{subsec:reddening}) since the relevant extinction maps do not go further than 4\,kpc. By saying that the $G=19.35$\,mag source has a main sequence star that contributes $\approx90\%$ of the flux in the $G$-band (hence alone is approximately $G=19.5$\,mag) and the AM~CVn contributes $\approx10\%$ (alone approximately $G=22.0\,$mag), we can calculate the surface density of $G\leq 22.0$\,mag stars from the \textsc{TRILEGAL} model and the average number that fall within a generously-assigned 0.1\arcsec\ radius of a position. Using a Poisson distribution, we obtained a chance alignment probability in the field of V407~Vul of 0.02\% ($3.5\sigma$). This is additionally a maximum chance alignment probability, since the foreground/background AM~CVn/main sequence star would have to be moving in the same direction as the other to show consistency between 16\,yr apart measurements in the astrometric wobble. All evidence strongly points towards V407~Vul being a gravitationally-bound AM~CVn and main sequence star system, leading us to promote it as the first gravitational wave detectable source located inside a hierarchal triple star system.

A loose constraint can now be placed on the outer orbit of the triple star system. The apparent spatial separation in astronomical units is given in Table~\ref{tab:periodTriple}, found using our adopted source distance. The eccentricity between the inner binary and main sequence star is unknown and so is the orbital inclination, so the semi-major axis between the inner binary and the tertiary is unknown too. But, we can crudely take the assumption that, although unlikely to be the case \citep[e.g.][]{Moe2017}, the triple orbit is circularised ($\textrm{eccentricity}=0$) and the semi-major axis is equal to the present day sky separation. A one-to-one conversion factor between the projected sky separation and the semi-major axis is a mode when integrating over the full eccentricity distribution for a large population of multi-star systems \citep{Dupuy2011, ElBadry2018imprintsGaia}, so these assumptions give the most-probable semi-major axis of the triple. We can also use a mass of 0.9\,M$_\odot$ for the main sequence star (Section~\ref{subsec:SEDfitparameters}) and take masses of each star in the AM~CVn of $0.6\pm0.2$\,\(\textup{M}_\odot\) for the accretor and $0.25\pm0.05$\,\(\textup{M}_\odot\) for the donor. These masses and mass errors were chosen to reflect the allowed mass range due to the impact spot location \citep{Barros2007HMCnc} and give a total mass for the inner binary of $0.85\pm0.21$\,M$_\odot$. Inserting these numbers into Kepler's third law, we obtain orbital periods of $1290\pm130$\,yr and $760\pm120$\,yr for the $g_s$- and $r_s$-band astrometric wobble detections, respectively. This triple system is far closer than those found in cataclysmic variables within 1000\,pc that contain a resolved tertiary component in \textit{Gaia} \citep{Shariat2025}, as depicted in Fig.~\ref{fig:tripleCVs}. We emphasise that the purpose of estimating a period of the triple orbit and the separation between the inner AM~CVn and the tertiary is to inform of their order of magnitudes. Clearly, \textit{Gaia} or similar missions in the next century are unlikely to be able to resolve a 3-body solution, and we are unable to resolve the 3-body orbit with follow-up ground-based observations on human timescales\footnote{One could repeat the methodology of Section~\ref{sec:Astrometry} and search for changes in the amplitude of astrometric wobble in the RA/Dec directions in the future to better verify the triple nature of the source. However, with an approximately 10$^3$\,yr orbital period, this is extremely challenging to measure.}.

\begin{figure}
    \centering
    \includegraphics[width=8.2cm]{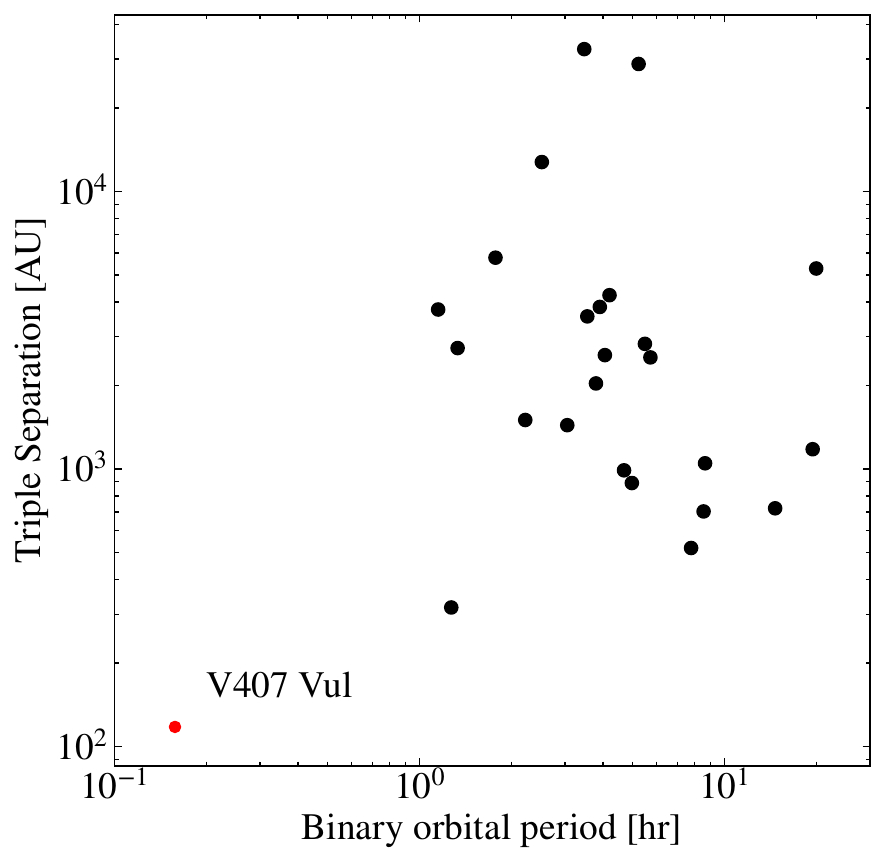}
    \caption{A comparison of the orbital period and separation of V407~Vul (red) and the cataclysmic variables that exist in a triple star systems presented in \citet{Shariat2025}.}
    \label{fig:tripleCVs}
\end{figure}

\subsection{Gravitational wave source}
\label{section:GravitationalWaveSource}
We have shown in this paper that V407~Vul is the only triple star system discovered to be detectable by millihertz-regime gravitational wave observatories. To calculate predictions for the Laser Interferometer Space Antenna (LISA), we take the source distance of 3510$^{+140}_{-110}$\,pc (Table~\ref{tab:fittedParameters}), a binary inclination of 60 degrees and masses $0.6\pm0.2$\,\(\textup{M}_\odot\) for the accretor WD and $0.25\pm0.05$\,\(\textup{M}_\odot\) 
 for the donor WD. We compute the characteristic strain with \citep{Shah2012, Moore2015, Kupfer2018LISAverificationBinaries}
\begin{equation}
     h_c = \frac{2(G\mathcal{M})^{5/3}}{c^4d}(\pi f_{GW})^{2/3}\sqrt{N_\textrm{cycle}}
\end{equation}
where $\mathcal{M}$ is the Chirp mass, $c$ the speed of light, $d$ the distance, $f_{GW}=2/P_\textrm{orb}$ the gravitational wave frequency and $N_\textrm{cycle}=fT_\textrm{obs}$. $T_\textrm{obs}$ is the observing time of LISA. 4\,yr is used as a nominal LISA mission time and 10\,yr is considered for the extended lifetime. The predicted characteristic strain after these times is $(3.97\pm1.35)\times10^{-20}$ and $(6.31\pm2.07)\times10^{-20}$, respectively.

Moreover, we can predict the signal strength of V407~Vul after these mission times using the \textsc{legwork} python package \citep{LEGWORK_apjs, LEGWORK_joss}, sampling with the same constraints and errors as before. We obtain a signal-to-noise ratio of $27.4\pm9.1$ after a 4\,yr \textit{LISA} mission time and $43.6\pm14.1$ after 10\,yr. These measurements are significantly lower than those that have been presented in past papers, but we emphasise that this stems from the somewhat unknown distance in the past that is now well constrained by our spectro-photometric fitting and new UV observations (Section~\ref{sec:SEDfittingSpectroscopicPhotometric}). V407~Vul thus remains one of the brightest ``verification binaries'' detectable with LISA. The $\approx10^3$\,yr orbit of the triple architecture is too large to impart any observable gravitational wave Doppler shift by LISA or other upcoming detectors, as this would need to be under ten times the gravitational wave detector mission time \citep{Robson2018}.

Gravitational wave detectable binaries inside triple star systems have been long anticipated, with none amongst the roughly 55 ``verification binaries'' \citep{Kupfer2024lisa}. A third star has the potential to influence the orbit of the inner binary through the von Zeipel–Kozai–Lidov mechanism \citep{vonZeipel1910, Lidov1962, Kozai1962} or other dynamical instabilities. \citet{Rajamuthukumar2025} predict that around 57\% of DWDs formed in a triple star system retain the tertiary companion once entering the LISA frequency band (sensitive to orbital periods of a couple of minutes to about 1.5\,hr). They find that another approximately 27\% lose the tertiary object before evolving into the LISA band, while about 9\% of gravitational wave detectable DWDs are only observable in this frequency space thanks to the presence of the tertiary, otherwise having merger times longer than a Hubble time if the binary were isolated.

Synthetic population models therefore predict many LISA detectable binaries to be found in triple systems harbouring a main sequence star, so V407~Vul being the only such case is surprising. The one other potential triple candidate hosting an outbursting AM~CVn together with a K-type star that outshines the source is MOA~2010-BLG-087 \citep{Green2020}. Its distance is loosely constrained so the source may not be gravitational wave detectable and the K-type star may not be gravitationally bound to the binary here. Sources like these are more observationally challenging in the optical since the binary is outshone, and hence the selection criteria used to discover ultra-compact binaries are not applicable \citep[e.g.][]{Burdge2020systematic, vanRoestel2021}. Upcoming \textit{Gaia} time-series astrometry and gravitational wave detections will be invaluable tools to discover new ultra-compact objects inside triple star systems. The other means of discovery will be with UV space observatories that separate the light of the hot, blue source from that of the main sequence companion, such as the Ultraviolet Explorer \citep[UVEX,][]{Kulkarni2021uves}, or from X-ray luminous sources \citep{2024PASP..136e4201R, 2025PASP..137a4201R}.

\subsection{The fate of V407 Vul}
\label{sec:FateOfV407Vul}
The final state of V407~Vul is unclear due to our lack of understanding of which exact conditions are required for a type Ia supernova, a merger or an outspiraling AM~CVn in a DWD binary evolution \citep{Munday2025esoWhitePaper}. The probability of a multi-star orbit surviving is very low in the case of massive stars due to the orbital instability caused by supernova explosions \citep{Pijloo2012, Lu2019, Toonen2020triple, Preece2024}. A very similar situation could occur for V407~Vul if the AM~CVn accretor accumulates a critical amount of helium during the full mass transfer phase and triggers a type Ia supernova via a detonation process \citep{Bildsten2006, Wong2023}. In this series of events, even with a surviving donor star, the tertiary object will be largely unaffected in its evolution and will become gravitationally unbound. The donor may be either destroyed in the supernova or ejected as a hyper-velocity star \citep[e.g.][]{Shen2018, Wong2024, Wong2025}

If dynamical instability causes the binary to merge without any sub-Chandrasekhar mass explosion, a new rejuvenated star will be born and the merger remnant will retain almost all of the mass of the WD ancestors \citep[e.g.][]{Shiber2024, Frank2025}. Mass limits based on the direct impact spot \citep{Barros2007HMCnc} indicate that a carbon-oxygen core WD accretor will combine with the helium core WD donor. Therefore, a new 0.6--1.1\,M$_\odot$ carbon-oxygen core WD would be the expected remnant after passing through a R~Coronae~Borealis and Extreme Helium star phase \citep{Webbink1984, IbenTutukov1984, Clayton2007, Wu2022}. There will be no change in the mass ratio between the merged binary and the main sequence star, so the main sequence star will be unaffected \citep[see also][]{Rao2012}.

Lastly, if the inner binary instead evolves to become an out-spiraling AM~CVn, the helium-core WD donor will be significantly stripped while the accretor accumulates mass to become a more massive carbon-oxygen WD. Assuming a close to unity mass retention factor, there will again be no difference in mass ratio between the inner binary and the tertiary object, so the triple dynamics will become largely unaltered. 

The main sequence tertiary will eventually evolve and expand before losing its envelope. For these cases where the main sequence star remains bound with a merger remnant or a surviving binary, we can take a constant inner binary to tertiary mass ratio of one. Dealing with the situation where the apparent sky separation is the semi-major axis of the triple, the Roche lobe radius is approximately 10\,000\,R$_\odot$. This is too large for stable Roche lobe overflow or a common envelope event with the inner binary or merger remnant to occur, the envelope of the tertiary will be lost as a stellar wind, and we can expect that the objects evolve in near-isolated conditions.


\section{Conclusion}
We have presented continued observations of the system V407~Vul that provide new insight into multiple astrophysical parameters of the star system. Even though V407~Vul is unresolved in \textit{Gaia}, ground-based astrometry using HiPERCAM on the GTC probes below the \textit{Gaia} diffraction limit and reveals an astrometric wobble of the centre of light on the 569\,s photometric period (Fig.~\ref{fig:timings_hcamLCs}). The sky separation of the sources is approximately 0.03--0.04\arcsec, providing clear evidence that the source of photometric variability is separate from the main sequence star that dominates the optical flux. We also witness consistency with the astrometric wobble from data spread 16 years apart, which, when considering the proper motion of the main sequence star, rules out the possibility of a stationary background source to at least a $3.2\sigma$ level. The chance alignment probability of another star is also low (0.02\%), and this star would have to be moving in the same direction of V407~Vul with a similar proper motion, making the probability of this event even smaller. Hence, to a very high degree of certainty, the spatial separation detection confirms that V407~Vul is a gravitationally bound triple star system. The most likely system architecture from all evidence in the past and presented in this study is that the photometrically variable, 569\,s inner binary is a DWD AM~CVn gravitationally bound with an outer tertiary companion. V407~Vul is therefore detectable by mHz~gravitational wave observatories, effectively becoming the first ``verification triple''.

Twenty additional years of time-series optical photometry now precisely constrains the orbital decay of the AM~CVn to a 1\% level. Continued orbital timing is encouraged to probe a second derivative term in the orbital decay. Moreover, we have been able to isolate light from the AM~CVn with newly presented HST/WFC3 imaging, showing a clear flux excess from an approximately $58\,000$\,K accretor WD. Flux from the AM~CVn dominates below approximately 4000\AA, which is shown by a flux excess in our Keck/LRIS and LBT/MODS spectra. High-quality spectroscopy bluer than 3500\AA~is now strongly encouraged to search for emission lines in the UV and near-UV, where the AM~CVn contributes almost 100\% of the light, and explore the formation history of the donor \citep[e.g.][]{Chakraborty2024}.

Our spectro-photometric fitting places the tightest distance constraint on V407~Vul to date, particularly important for its role as a gravitational wave detector ``verification binary'', or now ``verification triple''. A distance of $3510^{+140}_{-110}$\,pc gives rise to a LISA signal-to-noise ratio of $27.4\pm9.1$ after a 4\,yr mission time or $43.6\pm14.1$\,yr after a 10\,yr mission time, still marking V407~Vul as one of the brightest gravitational wave sources. No imprint of the tertiary object will be recognisable in the gravitational wave signal detected by LISA.

The future of the inner AM~CVn binary remains unclear, with the possible evolutionary channels being a surviving, outspiraling AM~CVn binary, a sub-luminous type Ia supernova or a merger of the two stars. The outer tertiary however is too distant to have any strong influence on the final state of the AM~CVn. Hence, while the two are gravitationally bound, the two entities should continue to evolve in near-isolated conditions.

\begin{acknowledgements}
We thank Holly Preece, Max Pritzkuleit and Silvia Toonen for useful conversation on elements of this study. We also thank the anonymous referee for improving the quality of this manuscript. Based on observations made with the William Herschel Telescope and the Isaac Newton Telescope operated on the island of La Palma by the Isaac Newton Group of Telescopes in the Spanish Observatorio del Roque de los Muchachos of the Instituto de Astrof\'isica de Canarias. Based on observations made with the NASA/ESA Hubble Space Telescope, obtained at the Space Telescope Science Institute, which is operated by the Association of Universities for Research in Astronomy, Inc., under NASA contract NAS 5-26555. Based on observations made with the Gran Telescopio Canarias (GTC), installed at the Spanish Observatorio del Roque de los Muchachos of the Instituto de Astrofísica de Canarias, on the island of La Palma. The construction of HiPERCAM was funded by the European Research Council under the European Union's Seventh Framework Programme (FP/2007-2013) under ERC-2013-ADG Grant Agreement no. 340040 (HiPERCAM). The operation of ULTRACAM and HiPERCAM are funded by the United Kingdom’s Science and Technology Facilities Council under grant ST/Z000033/1. N.R. is supported by the Deutsche Forschungsgemeinschaft (DFG) through grant RE3915/2-1. N.M. is supported by the Deutsches Zentrum fur Luft- und Raumfahrt (DLR) through grant 50 OR 2315. VSD is supported by a Leverhulme Research Fellowship. NCS acknowledges support from the Science and Technology Facilities Council (STFC) grant ST/X001121/1. M.D.\ is supported by the Deutsches Zentrum fur Luft- und Raumfahrt (DLR) through grant 50OR2510. SGP acknowledges support by the Science and Technology Facilities Council (grant ST/B001174/1). For the purpose of open access, the authors has applied a creative commons attribution (CC BY) licence to any author accepted manuscript version arising.
\end{acknowledgements}

%
   \bibliographystyle{aa} 
   \bibliography{V407Vul} 
\begin{appendix}




\onecolumn
\section{Observing Log}
\label{tab:observingLog}
\begin{table}[ht]
    \caption{An observing log of all V407~Vul observations acquired by us with HiPERCAM, ULTRACAM, ULTRSASPEC and the INT/WFC. If not stated differently in the comments, the full observing block is assumed to have no clouds. Given the crowded field around V407~Vul, observations were taken in image seeing conditions of less than $\approx 1.8$\arcsec\ always, aiming for $<1.4$\arcsec. The duration represents the time on target after acquisition. MJD$_{\textrm{mid}}$ is the modified Julian date at the centre of the observing period. In the instrument column, UCAM, USPEC and HCAM are abbreviations for ULTRACAM, ULTRASPEC and HiPERCAM.}
    \centering
    \begin{tabular}{c c c c c c c c}
    \toprule\toprule
         Night & MJD$_\textrm{mid}$  & Filters & Telescope & Instrument & Cadence (s) & Duration (min) & Comments\\
         \midrule
2003-05-21  &  52781.2  &  u$^\prime$g$^\prime$i$^\prime$  &  WHT  &  UCAM   &  9.8  &  48.0  &  Seeing 1.0\arcsec  \\
2003-05-22  &  52782.2  &  u$^\prime$g$^\prime$i$^\prime$  &  WHT  &  UCAM   &  9.7  &  143.1  &  Seeing 1.0\arcsec  \\
2003-05-23  &  52783.2  &  u$^\prime$g$^\prime$i$^\prime$  &  WHT  &  UCAM   &  9.7  &  167.6  &  Seeing 1.1\arcsec  \\
2003-05-24  &  52784.2  &  u$^\prime$g$^\prime$i$^\prime$  &  WHT  &  UCAM   &  9.7  &  136.2  &  Seeing 1.2\arcsec  \\
2003-05-25  &  52785.1  &  u$^\prime$g$^\prime$i$^\prime$  &  WHT  &  UCAM   &  9.7  &  118.0  &  Seeing 1.1-1.4\arcsec  \\
2005-08-27  &  53610.0  &  u$^\prime$g$^\prime$r$^\prime$  &  WHT  &  UCAM   &  14.9  &  186.5  &  Seeing 1.4  \\
2005-08-28  &  53610.9  &  u$^\prime$g$^\prime$r$^\prime$  &  WHT  &  UCAM   &  14.9  &  84.0  &  Seeing 1.0\arcsec  \\
2005-08-30  &  53612.9  &  u$^\prime$g$^\prime$r$^\prime$  &  WHT  &  UCAM   &  14.9  &  184.4  &  Seeing 1.0\arcsec  \\
2005-08-31  &  53613.9  &  u$^\prime$g$^\prime$r$^\prime$  &  WHT  &  UCAM   &  14.9  &  126.2  &  Seeing 1.0-1.4\arcsec  \\
2005-09-01  &  53614.9  &  u$^\prime$g$^\prime$r$^\prime$  &  WHT  &  UCAM   &  14.9  &  131.9  &  Seeing 1.0  \\
2008-08-06  &  54685.1  &  u$^\prime$g$^\prime$i$^\prime$  &  WHT  &  UCAM   &  5.7  &  50.0  &  Seeing 1.0\arcsec  \\
2011-05-28  &  55710.3  &  u$^\prime$g$^\prime$r$^\prime$  &  NTT  &  UCAM   &  10.1  &  26.7  &  Seeing 1.2\arcsec, cloudy  \\
2011-05-30  &  55712.3  &  u$^\prime$g$^\prime$r$^\prime$  &  NTT  &  UCAM   &  9.9  &  84.1  &  Seeing 1.2\arcsec  \\
2012-10-08  &  56208.8  &  u$^\prime$g$^\prime$r$^\prime$  &  WHT  &  UCAM   &  10.0  &  97.1  &  Seeing 1.1\arcsec  \\
2012-10-10  &  56210.9  &  u$^\prime$g$^\prime$r$^\prime$  &  WHT  &  UCAM   &  8.0  &  62.2  &  Seeing 1.2\arcsec  \\
2013-08-03  &  56508.1  &  u$^\prime$g$^\prime$r$^\prime$  &  WHT  &  UCAM   &  9.0  &  207.8  &  Seeing 1.2\arcsec  \\
2015-05-21  &  57164.2  &  u$^\prime$g$^\prime$r$^\prime$  &  WHT  &  UCAM   &  15.0  &  72.8  &  Seeing 1.4\arcsec  \\
2018-04-13  &  58222.3  &  u$^\prime$g$^\prime$r$^\prime$  &  NTT  &  UCAM   &  15.0  &  64.0  &  Seeing 1.0\arcsec  \\
2018-05-19  &  58258.1  &  u$_s$~g$_s$~r$_s$~i$_s$~z$_s$  &  GTC  &  HCAM  &  9.2  &  48.7  &  Seeing 0.8\arcsec  \\
2019-06-06  &  58641.1  &  u$_s$~g$_s$~r$_s$~i$_s$~z$_s$  &  GTC  &  HCAM  &  9.2  &  23.9  &  Seeing 0.7\arcsec  \\
2021-06-16  &  59382.2  &  u$_s$~g$_s$~r$_s$~i$_s$~z$_s$  &  GTC  &  HCAM  &  9.3  &  102.0  &  Seeing 0.7-0.9\arcsec  \\
2023-03-02 & 60006.3 & g & INT & WFC & 25.0 & 80.6 & Seeing 1.0\arcsec \\
2023-10-08 & 60226.0 & u$_s~$g$_s~$r$_s$ & NTT & UCAM & 10.0 & 40.1 & Seeing 1.5\arcsec\\
2023-10-13 & 60231.0 & u$_s~$g$_s$~i$_s$ & NTT & UCAM & 10.0 & 23.4 & Seeing 1.8\arcsec\\
2025-04-26 & 60792.3 & u$_s~$g$_s~$i$_s$ & NTT & UCAM & 14.8 & 17.9 & Seeing 1.7\arcsec\\
2025-04-26 & 60792.4 & u$_s~$g$_s~$i$_s$ & NTT & UCAM & 12.4 & 67.9 & Seeing 1.4\arcsec \\
2026-04-17 & 61148.4 & u$_s~$g$_s~$i$_s$ & NTT & UCAM & 11.0 & 95.8 & Seeing 1.1\arcsec\\
         \bottomrule
    \end{tabular}
\end{table}
\newpage
\section{Wavelength dependence of timing solutions}
\label{Appendix:WavelengthDependenceTiming}
\begin{figure}[ht]
    \centering
    \includegraphics[width=0.5\columnwidth, keepaspectratio, clip, trim={0.25cm 0.25cm 0.25cm 0.25cm}]{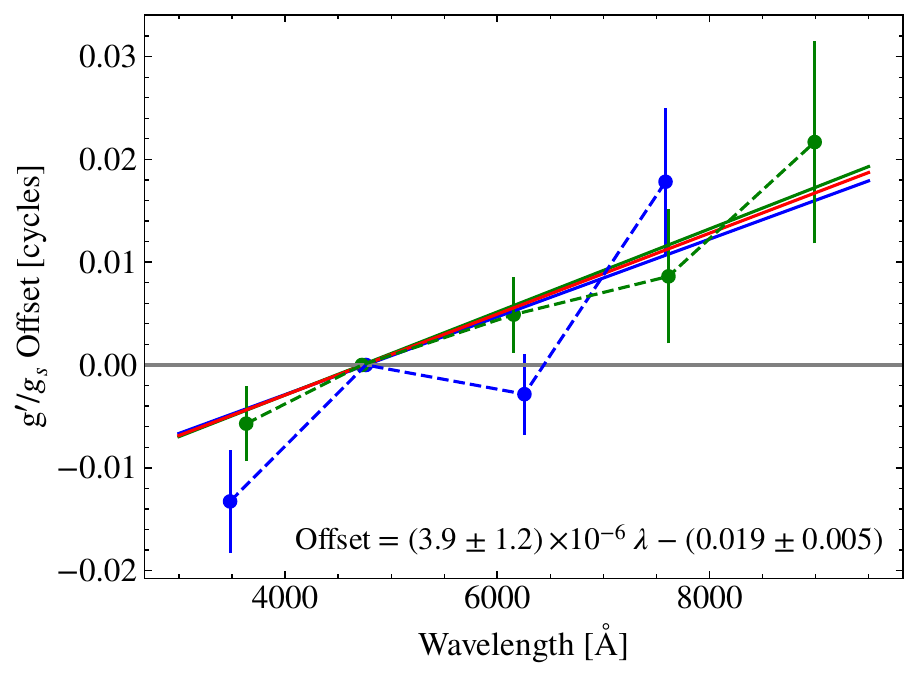}
    \caption{The wavelength dependence of timing solutions. Plotted is the relative timing offset from the $g^\prime$/$g_s$ band. The super SDSS $u_s$, $g_s$, $r_s$, $i_s$ and $z_s$ are plotted in green and the SDSS prime $u^\prime$, $g^\prime$, $r^\prime$ and $i^\prime$ filters in blue. The effective wavelength for each filter is assigned. The linear fits for each dataset are plotted in the respective colours, with the linear fit to all data combined in red. The linear fit to the combined data is annotated on the graph.}
    \label{fig:wavelengthDependence}
\end{figure}

\section{Absolute flux measurements}
\label{appendix:Photometry}
\begin{table}[ht]
    \label{tableappendix:Photometry}
    \caption{All flux calibrated photometry used in the spectro-photometric fitting. Filters beginning with ``Barros'' are SDSS prime measurements quoted in \citet{Barros2007HMCnc}. Other survey measurements are from the Sloan Digital Sky Survey \citep[SDSS][]{SDSSdr16} and Pan-STARRS \citep{Panstarrs}. The Pan-STARRS PS1:y filter was omitted since it showed a large and unusual excess. This likely originates from contamination with a neighbouring source, given the crowded nature of the field. Effective wavelengths ($\lambda_\textrm{eff}$) are listed as a reference point.}
    \centering
    \begin{tabular}{r c r r}
    \toprule\toprule
         Filter & $\lambda_\textrm{eff}$ [\AA]  & Flux [10$^{-28}$ erg/cm$^2$/s/Hz]\\
         \midrule
HST/F225W & 2373 & 0.1929  $\pm$  0.1062\\
XMM-OT/UVW1 & 2908 & 0.6000  $\pm$  0.1640\\
Barros/SDSSu$^\prime$ & 3482 & 0.8872  $\pm$  0.0654\\
HST/F390W & 4022 & 1.2221  $\pm$  0.1484\\
Barros/SDSSg$^\prime$ & 4762 & 2.7797  $\pm$  0.1536\\
PAN-STARRS/PS1:g & 4772 & 3.1400  $\pm$  0.0800\\
SDSS/g & 4820 & 3.1500  $\pm$  0.1000\\
PAN-STARRS/PS1:r & 6126 & 6.4800  $\pm$  0.0700\\
SDSS/r & 6247 & 6.5100  $\pm$  0.0700\\
Barros/SDSSr$^\prime$ & 6256 & 6.9183  $\pm$  0.6372\\
PAN-STARRS/PS1:i & 7480 & 10.2000  $\pm$  0.1000\\
Barros/SDSSi$^\prime$ & 7586 & 9.5499  $\pm$  0.4398\\
SDSS:i & 7635 & 10.2000  $\pm$  0.1000\\
PAN-STARRS/PS1:z & 8652 & 12.5000  $\pm$  0.2000\\
SDSS:z & 9018 & 12.5000  $\pm$  0.2000\\
         \bottomrule
    \end{tabular}
\end{table}

\clearpage
\section{Corner plot diagrams}
\label{appendix:Corners}
\begin{figure}[ht]
    \centering
    \includegraphics[width=\textwidth, clip, trim={0.1cm 0.2cm 0.6cm 0.25cm}]{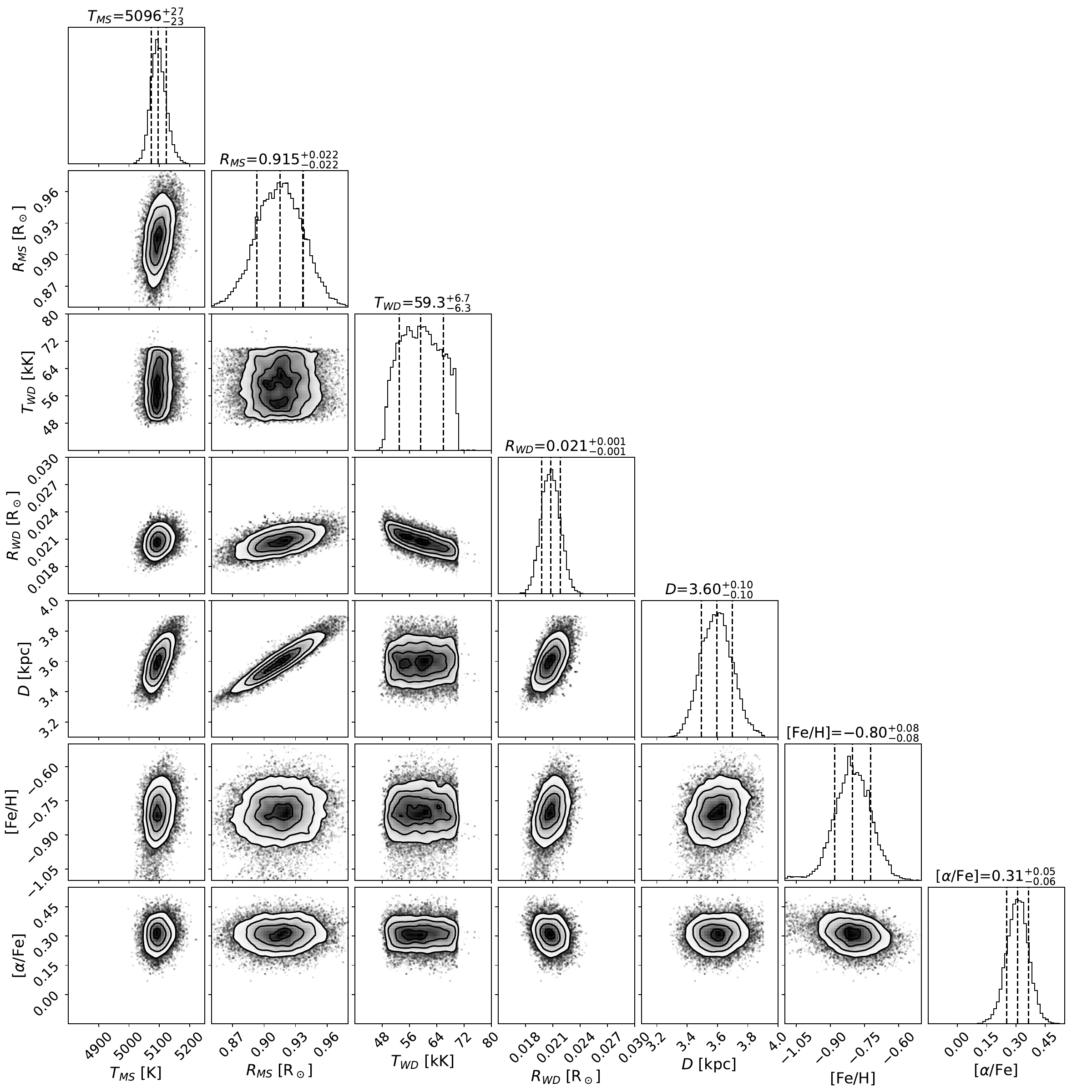}
    \caption{V407 Vul system parameters from spectro-photometric fitting with the Keck/LRIS spectrum. The slightly sudden upper limit of the posterior distribution for $T_\textrm{WD}$ is a result of an increasingly worse solution on approach to 70\,000\,K combined with the switch of synthetic spectral grids at this boundary (Section~\ref{subsec:spectrophotometricModelAtmospheres}). The donor WD is assumed to be of $T_\textrm{WD2}=8000$\,K and $R_\textrm{WD2}=0.035$\,R$_\odot$. Both WDs have the spectral shape for $\log(\textrm{g}) = 8.0$\,dex.}
    \label{fig:cornerKeck}
\end{figure}

\begin{figure}[ht]
    \centering
    \includegraphics[width=\textwidth, clip, trim={0.1cm 0.2cm 0.6cm 0.25cm}]{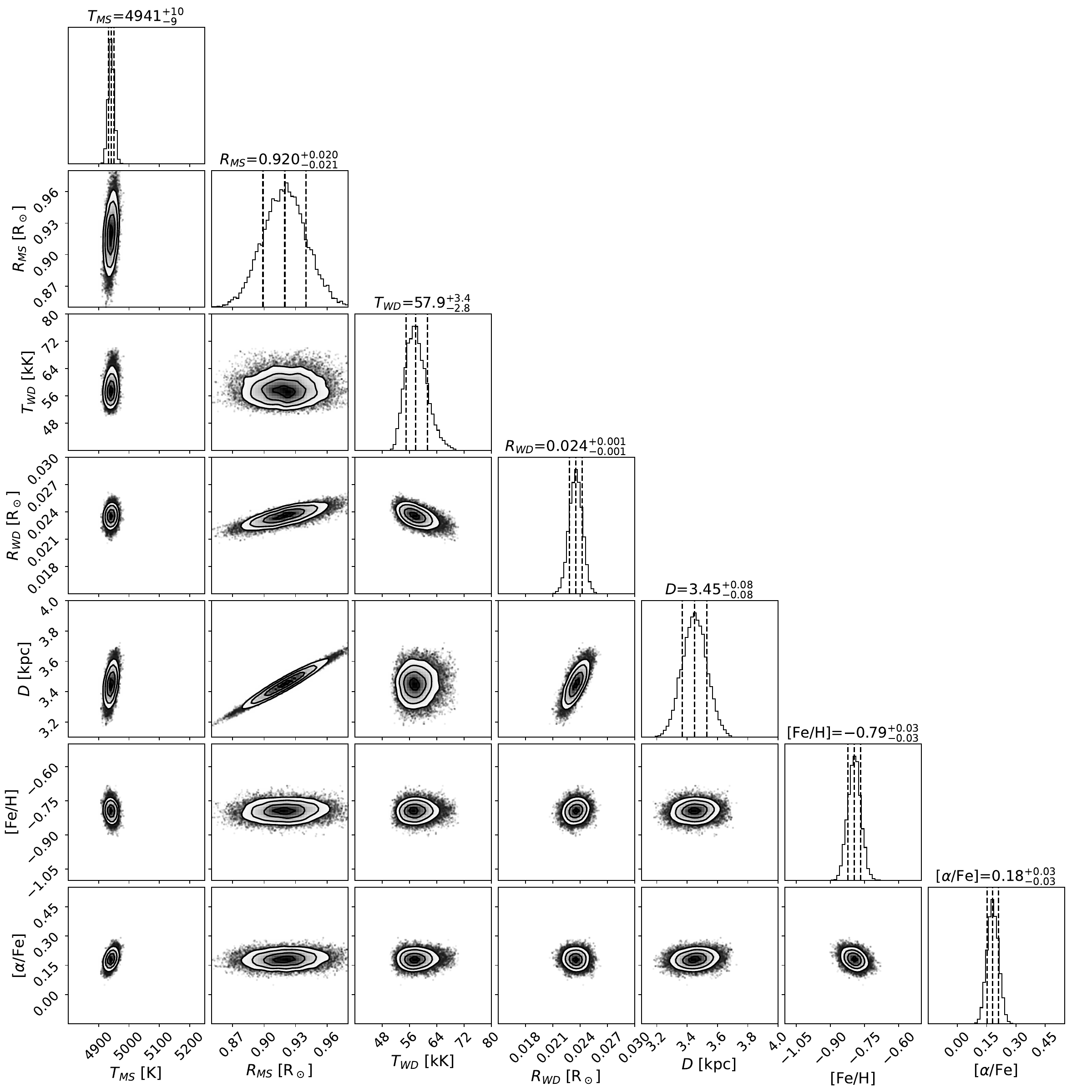}
    \caption{V407 Vul system parameters from spectro-photometric fitting with the LBT/MODS spectrum. The donor WD is assumed to be of $T_\textrm{WD2}=8000$\,K and $R_\textrm{WD2}=0.035$\,R$_\odot$. Both WDs have the spectral shape for $\log(\textrm{g}) = 8.0$\,dex.}
    \label{fig:cornerLBT}
\end{figure}
\end{appendix}
\end{document}